\documentclass[12pt]{article}
\usepackage{amssymb,amsmath,epsfig}
\begin{document} \parindent=0pt
\parskip=6pt \rm
\vspace*{0.5cm}
\begin{center}

{\bf \large Meissner superconductivity in itinerant ferromagnets}

\vspace*{0.5cm}

D. V. SHOPOVA, D. I. UZUNOV

CP Laboratory, G. Nadjakov Institute of Solid State Physics,\\
Bulgarian Academy of Sciences, BG-1784 Sofia, Bulgaria.
\end{center}

\begin{abstract} Recent results about the coexistence of ferromagnetism and
unconventional superconductivity with spin-triplet Cooper pairing
are reviewed on the basis of the quasi-phenomenological
Ginzburg-Landau theory. The superconductivity in the mixed phase
of coexistence of ferromagnetism and unconventional
superconductivity is triggered by the spontaneous magnetization.
The mixed phase is stable whereas the other superconducting phases
that usually exist in unconventional superconductors are either
unstable or metastable at relatively low temperatures in a quite
narrow domain of the phase diagram and the stability properties
are determined by the particular values of Landau parameters. The
phase transitions from the normal phase to the phase of
coexistence is of first order while the phase transition from the
ferromagnetic phase to the coexistence phase can be either of
first or second order depending on the concrete substance. The
Cooper pair and crystal anisotropy are relevant to a more precise
outline  of the phase diagram shape and reduce the degeneration of
the ground states of the system. The results are discussed in view
of application to itinerant ferromagnetic compounds as UGe$_2$,
ZrZn$_2$, URhGe.
\end{abstract}

\section[]{Introduction}

\subsection[]{Notes about unconventional superconductivity}

The phenomenon of unconventional Cooper pairing of fermions, i.e.,
the formation of Cooper pairs with nonzero angular momentum was
theoretically predicted~\cite{Pitaevskii:1959} in 1959 as a
mechanism of superfluidity in Fermi liquids. In 1972 the same
phenomenon - unconventional superfluidity due to a $p$-wave (spin
triplet) Cooper pairing of~$^3$He atoms, was experimentally
discovered in the mK range of temperatures; for details and
theoretical description see
Refs.~\cite{Leggett:1975,Vollhardt:1990,Volovik:2003}. Note that,
in contrast to the standard $s$-wave pairing in usual
(conventional) superconductors where the electron pairs are formed
by an attractive electron-electron interaction due to a virtual
phonon exchange, the widely accepted mechanism of the Cooper
pairing in superfluid $^3$He is based on an attractive interaction
between the fermions ($^3$He atoms) as a result of a virtual
exchange of spin fluctuations. Certain spin fluctuation mechanisms
of unconventional Cooper pairing of electrons are assumed also for
the discovered in 1979 heavy fermion superconductors (see, e.g.,
Refs.~\cite{Stewart:1984, Sigrist:1991, Mineev:1999}) as well as
for some classes
 of high-temperature superconductors (see, e.g., Refs.~\cite{Sigrist:1987,
Annett:1988, Volovik:1988, Blagoeva:1990, Uzunov:1990,
Uzunov:1993, Annett:1995, Harlingen:1995, Tsuei:2000}).

The possible superconducting phases in unconventional
superconductors are described in the framework of the general
Ginzburg-Landau (GL) effective free energy
functional~\cite{Uzunov:1993} with the help of the  symmetry
groups theory. Thus a variety of possible superconducting
orderings were predicted for different crystal
structures~\cite{Volovik0:1985, Volovik:1985, Ueda:1985,
Blount:1985, Ozaki:1985, Ozaki:1986}. A detailed thermodynamic
analysis~\cite{Blagoeva:1990, Volovik:1985} of the homogeneous
(Meissner) phases and a renormalization group
investigation~\cite{Blagoeva:1990} of the superconducting phase
transition up to the two-loop approximation have been also
performed (for a three-loop renormalization group analysis, see
Ref.~\cite{Antonenko:1994}; for effects of magnetic fluctuations
and disorder, see~\cite{Busiello:1991, Busiello:1990}). We shall
essentially use these results in our present consideration.

\subsection[]{Experimental predictions}

In 2000, experiments~\cite{Saxena:2000} at low temperatures ($T \sim 1$
K) and high pressure ($P\sim 1$ GPa) demonstrated the existence of spin
triplet superconducting states in the metallic compound UGe$_2$. The
superconductivity is triggered by the spontaneous magnetization of the
ferromagnetic phase that occurs at much higher temperatures. It
coexists with the superconducting phase in the whole domain of its
existence below $T \sim 1$ K; see also experiments from
Refs.~\cite{Huxley:2001, Tateiwa:2001}, and the discussion in
Ref.~\cite{Coleman:2000}. The same phenomenon of existence of
superconductivity at low temperatures and high pressure in the domain
of the $(T,P)$ phase diagram where the ferromagnetic order is present
was observed in other ferromagnetic metallic compounds
(ZrZn$_2$~\cite{Pfleiderer:2001} and URhGe~\cite{Aoki:2001}) soon after
the discovery~\cite{Saxena:2000} of superconductivity in UGe$_2$.

In superconducting ternary and Chevrel compounds the influence of
magnetic order on superconductivity is also substantial (see,
e.g.,~\cite{Vonsovsky:1982, Maple:1982, Sinha:1984, Kotani:1984})
but in the newly found ferromagnetic substances the phase
transition temperature ($T_f$) to the ferromagnetic state is much
higher than the phase transition temperature ($T_{FS})$ from
ferromagnetic to a mixed state of coexistence of ferromagnetism
and superconductivity. For example, in UGe$_2$, $T_{FS} = 0.8$ K
while the critical temperature of the phase transition from
paramagnetic to ferromagnetic state in the same material is $T_f
=35$ K~\cite{Saxena:2000, Huxley:2001}. It can be assumed that in
these substances
 the material parameter $T_s$ defined as the
usual critical temperature of the second order phase transition from
normal to uniform (Meissner) supercondicting state in a zero external
magnetic field is much lower than the phase transition temperature
$T_{FS}$. The above mentioned experiments on the compounds~UGe$_{2}$,
URhGe, and ZrZn$_2$ do not give any evidence for the existence of a
standard normal-to-superconducting phase transition in a zero external
magnetic field.

It seems that the superconductivity in the metallic compounds mentioned
above always coexists with the ferromagnetic order and is enhanced by
it. In these systems, as claimed in Ref.~\cite{Saxena:2000}, the
superconductivity probably arises from the same electrons that create
the band magnetism and can be most naturally understood rather as a
triplet than spin-singlet pairing phenomenon.  Metallic compounds
UGe$_{2}$, URhGe, and ZrZn$_2$, are itinerant ferromagnets. An
unconventional superconductivity is also suggested~\cite{Saxena:2001}
as a possible outcome of recent experiments in Fe~\cite{Shimizu:2001},
in which a superconducting phase has been  discovered at temperatures
below $2$ K and pressures between 15 and 30 GPa. There both vortex and
Meissner superconductivity phases~\cite{Shimizu:2001} are found in the
high-pressure crystal modification of Fe with a hexagonal close-packed
lattice for which the strong ferromagnetism of the usual bcc iron
crystal probably disappears~\cite{Saxena:2001}. It can be hardly
claimed that in hexagonal Fe the ferromagnetism and superconductivity
coexist  but the clear evidence for a superconductivity is also a
remarkable achievement.

The reasonable question whether these examples of superconductivity and
coexistence of superconductivity and ferromagnetism are bulk or surface
effects can be stated. The earlier experiments performed before 2004 do
not answer this question. Recent experiments~\cite{Yelland:2005} show
that surface superconductivity appears in ZrZn$_2$ and its presence
depends essentially on the way of preparation of the sample. But in our
study it is important that bulk superconductivity can be considered
well established in this substance.

\subsection[]{Ferromagnetism versus superconductivity}

The important point in all discussions of the interplay of
superconductivity and ferromagnetism is that a small amount of
magnetic impurities can destroy superconductivity in conventional
($s$-wave) superconductors by breaking up the ($s$-wave) electron
pairs with opposite spins (paramagnetic impurity
effect~\cite{Abrikosov:1960}). In this aspect the phenomenological
arguments~\cite{Ginzburg:1956} and the conclusions on the basis of
the microscopic theory of magnetic impurities in $s$-wave
superconductors~\cite{Abrikosov:1960} are in a complete agreement
with each other; see, e.g., Refs.~\cite{Vonsovsky:1982,
Maple:1982, Sinha:1984, Kotani:1984}. In fact, a total suppression
of conventional ($s$-wave) superconductivity should occur in the
presence of an uniform spontaneous magnetization
$\mbox{\boldmath$M$}$, i.e., in a standard ferromagnetic
phase~\cite{Ginzburg:1956}. The physical reason for this
suppression is the same as in the case of magnetic impurities,
namely, the opposite electron spins in the $s$-wave Cooper pair
turn over along the vector $\mbox{\boldmath$M$}$ in order to lower
their Zeeman energy and, hence, the pairs break down. Therefore,
the ferromagnetic order can hardly coexist with conventional
superconducting states. Especially, this is valid for the
coexistence of uniform superconducting and ferromagnetic states
where the superconducting order parameter
$\psi(\mbox{\boldmath$x$})$ and the magnetization
$\mbox{\boldmath$M$}$ do not depend on the spatial vector
$\mbox{\boldmath$x$}$.

But yet a coexistence of $s$-wave superconductivity and
ferromagnetism may appear in uncommon materials and under quite
special circumstances. Furthermore, let us emphasize that the
conditions for the coexistence of nonuniform (``vertex'',
``spiral'', ``spin-sinosoidal'' or
``helical''~\cite{Vonsovsky:1982, Maple:1982})
 superconducting and ferromagnetic states are
 less restrictive than those for the
coexistence of uniform superconducting and ferromagnetic orders.
Coexistence of nonuniform phases has been discussed in details,
both experiment and theory, in ternary and Chevrel-phase compounds
where such a coexistence seems quite likely; for a comprehensive
review, see, for example, Refs. ~\cite{Vonsovsky:1982, Maple:1982,
Sinha:1984,Kotani:1984, Buzdin:1983}.

In fact the only two superconducting systems for which the
experimental data allow assumptions in a favor of a coexistence of
superconductivity and ferromagnetism are the rare earth ternary
boride compound ErRh$_4$B$_4$ and the Chervel phase compound
HoMo$_6$S$_8$; for a more extended review, see
Refs.~\cite{Maple:1982, Machida:1984}. In these compounds the
phase of coexistence appears in a very narrow temperature region
just below the Curie temperature $T_f$ of the ferromagnetic phase
transition. At lower temperatures the magnetic moments of the rare
earth 4$f$ electrons become better aligned, the magnetization
increases and the $s$-wave superconductivity pairs formed by the
conduction electrons disintegrate.

\subsection[]{Unconventional superconductivity triggered by
ferromagnetic order}

We shall not consider all important aspects of the long standing
problem of coexistence of superconductivity and ferromagnetism
rather we shall concentrate our attention on the description of
the newly discovered coexistence of ferromagnetism and
unconventional (spin-triplet) superconductivity in the itinerant
ferromagnets UGe$_2$, ZrZn$_2$, and URhGe. Here we wish to
emphasize that the main object of our discussion is the
superconductivity of these compounds and at a second place in the
rate of importance we put the problem of coexistence. The reason
is that the existence of superconductivity in itinerant
ferromagnets is a highly nontrivial phenomenon. As noted in
Ref.~\cite {Machida:2001} the superconductivity in these materials
appears to be difficult to explain in terms of previous
theories~\cite {Vonsovsky:1982, Maple:1982, Kotani:1984} and
 requires new concepts to interpret the experimental data.

We have already mentioned that in ternary compounds the
ferromagtetism comes from the localized 4$f$ electrons while the
s-wave Cooper pairs are formed by conduction electrons. In UGe$_2$
and URhGe the 5$f$ electrons of U atoms form both
superconductivity and ferromagnetic order~\cite{Saxena:2000,
Aoki:2001}. In ZrZn$_2$ the same
 double role is played by the 4$d$ electrons of Zr.
Therefore, the task is to describe this behavior of the band
electrons at a microscopic level. One may speculate about a
spin-fluctuation mediated unconventional Cooper pairing as is in
case of $^3$He and heavy fermion superconductors. These important
issues have not yet a reliable answer and for this reason we shall
confine our consideration to a phenomenological level.

In fact, a number of reliable experimental data as the coherence
length and the superconducting gap measurements~\cite{Saxena:2000,
Huxley:2001, Aoki:2001, Pfleiderer:2001} are in favor of the
conclusion about a spin-triplet Cooper pairing in these metallic
compounds, although the mechanism of pairing remains unclear. We
shall essentially use this reliable conclusion. This point of view
is consistent with the experimental observation of coexistence of
superconductivity only in a low temperature part of the
ferromagnetic domain of the phase diagram ($T,P$), which means
that a pure (non-ferromagnetic) superconducting phase is not
observed, a circumstance, that is also in favor of the assumption
of a spin-triplet superconductivity.  Our investigation leads to
results which confirm this general picture.

 On the basis of the experimental data
and conclusions presented for the first time in
Refs.~\cite{Saxena:2000, Coleman:2000} and shortly afterwards
confirmed in Refs.~\cite{Huxley:2001, Tateiwa:2001,
Pfleiderer:2001, Aoki:2001} one may reliably accept that the
superconductivity in these magnetic compounds is considerably
enhanced by the ferromagnetic order parameter
$\mbox{\boldmath$M$}$ and, perhaps, it could not exist without
this ``mechanism of ferromagnetic trigger,'' or, in short,
``$\mbox{\boldmath$M$}$-trigger''; see Refs.~\cite{Shopova1:2003,
 Shopova2:2003, Shopova1:2005} where this concept has been introduced for the first time.
  The trigger phenomenon is possible
for spin-triplet Cooper pairs where the electron spins point
parallel to each other and their turn along the vector of the
spontaneous magnetization $\mbox{\boldmath$M$}$ does not produce a
 breakdown of the spin-triplet Cooper pairs but rather stabilizes them and,
perhaps, stimulates their creation. We shall describe this
phenomenon at a phenomenological level.

\subsection[]{Phenomenological studies}

A phenomenological theory that explains the coexistence of
ferromagnetism and unconventional spin-triplet superconductivity
of GL type has been developed recently in ~\cite{Machida:2001,
Walker:2002} where possible low-order couplings between the
superconducting and ferromagnetic order parameters are derived
with the help of general symmetry group arguments. On this basis
several important features of the superconducting vortex state of
unconventional ferromagnetic superconductors were
demonstrated~\cite{Machida:2001, Walker:2002}.

In our review we shall follow the approach from
Refs.~\cite{Machida:2001, Walker:2002}  to investigate the conditions
for the occurrence of the Meissner phase and to demonstrate that the
presence of ferromagnetic order enhances the $p$-wave
superconductivity. We also establish the phase diagram of
 ferromagnetic superconductors in a zero
external magnetic field and show that the phase transition to the
superconducting state  can be either of first or second order depending
on the particular substance. We confirm the predictions made in
Refs.~\cite{Machida:2001,Walker:2002} about the symmetry of the ordered
phases.

In our study we use the mean-field
approximation~\cite{Uzunov:1993} and known results about the
possible phases in nonmagnetic superconductors with triplet
($p$-wave) pairing~\cite{Volovik:1985, Blagoeva:1990, Uzunov:1990,
Sigrist:1991}. Our results~\cite{Shopova1:2003, Shopova2:2003,
Shopova3:2003} show that taking into account the anisotropy of the
spin-triplet Cooper pairs modifies but does not drastically change
the thermodynamic properties of the coexistence phase, especially
in the temperature domain above the superconducting critical
temperature $T_s$. The effect  of crystal anisotropy is similar
but we shall not make an overall thermodynamic analysis of this
problem  because we have to consider  concrete systems and crystal
structures~\cite{Volovik:1985, Sigrist:1991} for which there is no
enough information from experiment to make conclusions about the
parameters of the theory. Our results confirm the general concept
that the anisotropy reduces the degree of ground state
degeneration, and depending on the symmetry of the crystal, picks
up a crystal direction for the ordering.

There exists a formal similarity between the phase diagram we
obtain and the phase diagram of certain improper
ferroelectrics~\cite{Gufan:1980, Gufan:1981, Latush:1985,
Toledano:1987, Gufan:1982, Cowley:1980}. We shall make use of the
concept in  the theory of improper ferroelectrics, where the
trigger of the primary order parameter  by a secondary order
parameter (the electric polarization) has been initially
introduced and exploited; see Ref.~\cite{Toledano:1987,
Gufan:1982, Cowley:1980}. The mechanism of the M-triggered
superconductivity in itinerant ferromagnets~\cite{Shopova1:2003,
Shopova2:2003, Shopova1:2005} is formally identical to the
mechanism of appearance of structural order triggered by the
electric polarization in improper ferroelectrics (see, e.g.,
Refs.~\cite{Toledano:1987, Gufan:1982, Cowley:1980}).

Our investigation is based on the GL free energy functional of
unconventional ferromagnetic superconductors presented in Sec.~2.1
and we shall establish the uniform phases which are described by
it. More information about the justification of this investigation
is presented in Sec.~2.2. We work with a quite general GL free
energy and the problem is that there is no enough information
about the values of the parameters of the model for concrete
compounds (UGe$_2$, URhGe, ZrZn$_2$) where the ferromagnetic
superconductivity has been discovered. On the one hand the lack of
information makes impossible a detailed comparison of the theory
to the available experimental data but on the other hand our
results are not bound to one or more concrete substances and can
be applied to any unconventional ferromagnetic superconductor. In
Sec.~3 we discuss the phases in nonmagnetic unconventional
superconductors. In Sec.~4 the M-trigger
effect~\cite{Shopova1:2003, Shopova2:2003, Shopova1:2005}
 will be described when only a linear coupling  of the magnetization
 $\mbox{\boldmath$M$}$
to the superconducting order parameter $\psi$ is considered; here
the spatial dependence of order parameters and all anisotropy
effects are ignored. In Sec.~5 we analyze the influence of
quadratic coupling of magnetization to the superconducting order
parameter on the thermodynamics of the ferromagnetic
superconductors. The application of our results to experimental
$(T,P)$ phase diagrams is discussed in Sec. 5.3. In Sec.~6
 the anisotropy effects are outlined. Our main attention is focussed on the
 Cooper-pair anisotropy. Note, that certain types of crystal anisotropy may produce
more than one ferromagnetic phase~\cite{Kimura:2004, Uhlarz:2004,
Aso:2005} but here we shall not dwell on this interesting topic.
In Sec.~7 we summarize our conclusions.

\section[]{Ginzburg-Landau free energy}

Following Refs.~\cite{Volovik:1985, Machida:2001, Shopova1:2003, Shopova2:2003,
 Shopova1:2005, Walker:2002} in
this Chapter we discuss the phenomenological theory of
spin-triplet ferromagnetic superconductors and justify our
consideration in Sections 3--6.

\subsection[]{Model}

The general GL free energy functional, we shall use in our analysis, is
\begin{equation}
\label{eq1} F[\psi,\mbox{\boldmath$M$}]=\int d^3 x f(\psi,
\mbox{\boldmath$M$}),
\end{equation}
where the free energy density $f(\psi,\mbox{\boldmath$M$})$ ( hereafter
called ``free energy'') of a spin-triplet ferromagnetic superconductor
is a sum of five terms~\cite{Machida:2001, Walker:2002, Volovik:1985},
namely,
\begin{equation}
\label{eq2} f(\psi, \mbox{\boldmath$M$}) = f_{\mbox{\scriptsize
S}}(\psi) + f^{\prime}_{\mbox{\scriptsize F}}(\mbox{\boldmath$M$})
+ f_{\mbox{\scriptsize I}}(\psi,\mbox{\boldmath$M$}) +
\frac{\mbox{\boldmath$B$}^2}{8\pi} - \mbox{\boldmath$B.M$}.
\end{equation}
In Eq.~(2) the three dimensional complex vector $\psi =
\left\{\psi_j;j=1,2,3\right\}$ represents the superconducting
order parameter, $\mbox{\boldmath$B$} = (\mbox{\boldmath$H$} +
4\pi\mbox{\boldmath$M$}) = \nabla \times \mbox{\boldmath$A$}$ is
the magnetic induction; $\mbox{\boldmath$H$}$ is the external
magnetic field, $\mbox{\boldmath$A$} = \left\{A_j;
j=1,2,3\right\}$ is the magnetic vector potential. The last two
terms on  r.h.s. of Eq.~(2) are related with the magnetic energy
which includes both diamagnetic and paramagnetic effects in the
superconductor; see, e.g., \cite{Vonsovsky:1982, Blount:1979}.

The energy part $f_{\mbox{\scriptsize S}}(\psi)$ in Eq.~(2)
describes the superconductivity for $\mbox{\boldmath$H$} =
\mbox{\boldmath$M$} \equiv 0$. It can  be written in the form
\begin{equation}
\label{eq3} f_{\mbox{\scriptsize S}}(\psi)= f_{grad}(\psi)
 + a_s|\psi|^2 +\frac{b_s}{2}|\psi|^4 + \frac{u_s}{2}|\psi^2|^2 +
\frac{v_s}{2}\sum_{j=1}^{3}|\psi_j|^4.
\end{equation}
Here
\begin{eqnarray}
\label{eq4} f_{grad}(\psi)& = & K_1(D_i\psi_j)^{\ast}(D_i\psi_j)+K_2 [
 (D_i\psi_i)^{\ast}(D_j\psi_j)\\ \nonumber
 &&  +  (D_i\psi_j)^{\ast}(D_j\psi_i) ]
+ K_3(D_i\psi_i)^{\ast}(D_i\psi_i),
\end{eqnarray}
where a summation over the indices $i,j=1,2,3$ is assumed and the
symbol
\begin{equation}
\label{eq5}
 D_j = - i\hbar\frac{\partial}{\partial x_i} + \frac{2|e|}{c}A_j
\end{equation}
denotes a covariant differentiation. In Eq.~(3), $b_s
> 0$ and $a_s = \alpha_s(T-T_s)$, where $\alpha_s$ is a positive
material parameter and $T_s$ is the critical temperature of the
standard second order phase transition which may occur at $H =
{\cal{M}} = 0$; $H =|\mbox{\boldmath$H$}|$, and ${\cal{M}} =
|\mbox{\boldmath$M$}|$. The quantities $u_s$ and $v_s$ describe
the anisotropy of the spin-triplet Cooper pair  and the crystal
anisotropy, respectively, ~\cite{Volovik:1985,Blagoeva:1990}.
  Parameters $K_j$, $(j = 1,2,3)$ in Eq.~(4) are related with the
effective mass tensor of anisotropic Cooper pairs~\cite{Volovik:1985}.

The superconducting part (3) of the free energy $f(\psi, M)$ is derived
from symmetry group arguments and is independent of particular
microscopic models; see, e.g., Refs.~\cite{Volovik:1985, Sigrist:1991}.
According to classifications~\cite{Volovik:1985, Sigrist:1991} the
$p$-wave superconductivity in the cubic point group $O_h$ can be
realized through one-, two-, and three-dimensional representations of
the order parameter. The expressions (3) and (5) incorporate all three
possible cases. The coefficients $b_s$, $u_s$, and $v_s$ in Eq.~(3) are
different for weak and strong spin-orbit couplings but in our
investigation they are considered as undetermined material parameters
which depend on the particular substance.

The  free energy of a standard isotropic ferromagnet is given by
the term $f^{\prime}_{\mbox{\scriptsize F}}(\mbox{\boldmath$M$})$
in Eq.~(2),
\begin{equation}
\label{eq6} f^{\prime}_{\mbox{\scriptsize F}}(\mbox{\boldmath$M$}) =
c_f\sum_{j=1}^{3}|\nabla_j\mbox{\boldmath$M$}_j|^2 +
 a_f(T^{\prime}_f)\mbox{\boldmath$M$}^2 +
 \frac{b_f}{2}\mbox{\boldmath$M$}^4,
\end{equation}
where $\nabla_j = \partial/\partial x_j$ and $b_f > 0$. The
quantity $a_f(T^{\prime}_f) = \alpha_f(T-T^{\prime}_f)$ is
expressed by the material parameter $\alpha_f > 0$ and the
temperature $T^{\prime}_f$ which is different from the critical
temperature $T_f$ of the ferromagnet and this point will be
discussed below. We have already added a negative term ($-2\pi
{\cal{M}}^2$) to the total free energy
$f(\psi,\mbox{\boldmath$M$})$ and that is obvious by setting $H =
0$ in Eq.~(2). The negative energy ($-2\pi{\cal{M}}^2$) should be
added to $f^{\prime}_{\mbox{\scriptsize F}}(\mbox{\boldmath$M$})$.
In this way one obtains the total free energy
$f_{\mbox{\scriptsize F}} (\mbox{\boldmath$M$})$ of the
ferromagnet in a zero external magnetic field that is given by a
modification of Eq.~(6) according to the rule
\begin{equation}
\label{eq7} f_{\mbox{\scriptsize F}} (a_f) =
f^{\prime}_{\mbox{\scriptsize F}} \left[a_f(T^{\prime}_f) \rightarrow
a_f(T_f) \right],
\end{equation}
where  $a_f = \alpha_f (T - T_f)$ and
\begin{equation}
\label{eq8} T_f =  T^{\prime}_f + \frac{2\pi}{\alpha_f}
\end{equation}
is the critical temperature of a standard ferromagnetic phase
transition of second order. This scheme was used in studies of
rare earth ternary compounds~\cite{Vonsovsky:1982, Blount:1979,
Greenside:1980, Ng:1997}. Alternatively~\cite{Kuper:1980}, one may
use from the beginning the total ferromagnetic free energy
$f_{\mbox{\scriptsize F}}(a_f,\mbox{\boldmath$M$})$ as given by
Eqs.~(6)~-~(8) but in this case the magnetic energy included in
the last two terms on r.h.s. of Eq.~(2) should be replaced with
$H^2/8\pi$. Both approaches are equivalent.

The interaction between the ferromagnetic order parameter
$\mbox{\boldmath$M$}$ and the superconducting order parameter
$\psi$ is given by~\cite{Machida:2001,Walker:2002}
\begin{equation}
\label{eq9} f_{\mbox{\scriptsize I}}(\psi, \mbox{\boldmath$M$}) =
i\gamma_0 \mbox{\boldmath$M$}.(\psi\times \psi^*) + \delta
\mbox{\boldmath$M$}^2 |\psi|^2.
\end{equation}
The $\gamma_0$-term in the above expression is the most
substantial for the description of experimentally found
ferromagnetic superconductors~\cite{Walker:2002} and the $\delta
\mbox{\boldmath$M$}^2 |\psi|^2$--term makes the model more
realistic in the strong coupling limit as it gives the opportunity
to enlarge the phase diagram including both positive and negative
values of the parameter $a_s$. In this way  the domain of the
stable ferromagnetic order is extended down to zero temperatures
for a wide range of values of  material parameters and the
pressure $P$, a situation that corresponds to the experiments in
ferromagnetic superconductors.

In Eq.~(9) the coupling constant $\gamma_0 >0$ can be represented in
the form $\gamma_0 = 4\pi J$, where $J > 0$ is the ferromagnetic
exchange parameter~\cite{Walker:2002}. In general, the parameter
$\delta$ for ferromagnetic superconductors may take
 both positive and negative values. The values of the material parameters
($T_s$, $T_f$, $\alpha_s$, $\alpha_f$, $b_s$, $u_s$, $v_s$, $b_f$,
$K_j$, $\gamma_0$ and $\delta$) depend on the choice of the concrete
substance and on thermodynamic parameters as temperature $T$ and
pressure $P$.

\subsection[]{Way of treatment}

The total free energy (2) is a quite complex object of theoretical
investigation. The possible vortex and uniform phases of the model
cannot be investigated within a single calculation, rather one
should focus on concrete problems. In Ref.~\cite{Walker:2002} the
vortex phase was discussed with the help of the
criterion~\cite{Abrikosov:1957} for a stability of this state near
the phase transition line $T_{c2}(H)$; see also,
Ref.~\cite{Lifshitz:1980}. In case of $H = 0$ one should apply the
same criterion with respect to the magnetization ${\cal{M}}$ for
small values of $|\psi|$ near the phase transition line
$T_{c2}({\cal{M}})$ as performed in Ref.~\cite{Walker:2002}.

Here we shall be interested in the uniform phases when the order
parameters $\psi$ and $\mbox{\boldmath$M$}$ do not depend on the
spatial vector $\mbox{\boldmath$x$}\in V$, where $V$ is the volume
of the superconductor. We shall restrict our analysis to the
consideration of the coexistence of uniform (Meissner) phases and
ferromagnetic order. We shall make a detailed investigation in
order to show that the main properties of the uniform phases can
be well determined within an approximation when the crystal
anisotropy is neglected. Even, some of the main features of the
uniform phases in unconventional ferromagnetic superconductors can
be reliably outlined when the Cooper pair anisotropy is neglected,
too.

The assumption of a uniform magnetization $\mbox{\boldmath$M$}$ is
always reliable outside a quite close vicinity of the magnetic
phase transition and under the condition that the superconducting
order parameter $\psi$ is also uniform, i.e. that vortex phases
are not present at the respective temperature domain. This
conditions are directly satisfied in type I superconductors but in
type II superconductors the temperature should be sufficiently low
and the external magnetic field should be zero. Nevertheless, the
mentioned conditions for type II superconductors may turn
insufficient for the appearance of uniform superconducting states
in materials with quite high values of the spontaneous
magnetization. In such situation the uniform (Meissner)
superconductivity and, hence, the coexistence of this
superconductivity with uniform ferromagnetic order may not appear
even at zero temperature. Up to now type I unconventional
ferromagnetic superconductors have not been found whereas the
experimental data for the recently discovered compounds UGe$_2$,
URhGe, and ZrZn$_2$ are not enough to conclude definitely either
about the lack or the existence of uniform superconducting states
at low and ultra-low temperatures.

If real materials can be modelled by the general GL free energy
(1)~-~(9), the ground state properties will be described by
uniform states, which we shall investigate. The problem about the
availability of such states in real materials at finite
temperatures is quite subtle at the present stage of research when
the experimental data are not enough. Recently in
Ref.~\cite{Kotegawa:2004} an experimental evidence was given for
the coexistence of uniform superconductivity and ferromagnetism in
UGe$_2$. Thus we shall assume that uniform phases may exist in
some unconventional ferromagnetic superconductors. Moreover, we
have to emphasize that these phases appear as solutions of the GL
equations corresponding to the free energy (1)~-~(9). These
arguments completely justify our study.

In case of a strong easy axis type of magnetic anisotropy, as is
in UGe$_2$~\cite{Saxena:2000}, the overall complexity of
mean-field analysis of the free energy $f(\psi,
\mbox{\boldmath$M$})$ can be avoided by performing an
``Ising-like'' description: $\mbox{\boldmath$M$} =
(0,0,{\cal{M}})$, where ${\cal{M}} = \pm |\mbox{\boldmath$M$}|$ is
the magnetization along the ``$z$-axis.'' Because of the
equivalence of the ``up'' and ``down'' physical states $(\pm
\mbox{\boldmath$M$})$ the thermodynamic analysis can be performed
within the ``gauge" ${\cal{M}} \geq 0$. When the magnetic order
has a continuous symmetry we can take advantage of the symmetry of
the total free energy $f(\psi, \mbox{\boldmath$M$})$ and avoid the
consideration of equivalent thermodynamic states that occur as a
result of the respective symmetry breaking at the phase transition
point but have no effect on thermodynamics of the system. In the
isotropic system one may again choose a gauge, in which the
magnetization vector has the same direction as  $z$-axis
($|\mbox{\boldmath$M$}| = M_z = {\cal{M}}$) and this will not
influence the generality of thermodynamic analysis. Here we shall
prefer the alternative description within which the ferromagnetic
state may appear through two equivalent {\em up} and {\em down}
domains with magnetizations $ {\cal{M}}$ and ($ -{\cal{M}}$),
respectively.

We shall make the mean-field analysis of the uniform phases and
the possible phase transitions between such phases in a zero
external magnetic field ($\mbox{\boldmath$H$}=0)$, when the
crystal anisotropy is neglected ($v_s \equiv 0$). The only
exception will be the consideration
 in Sec.~3, where we briefly discuss the nonmagnetic superconductors
(${\cal{M}} \neq 0$).

We shall use notations in which the number of
 parameters is reduced. For this reason we introduce
\begin{equation}
\label{eq10} b = (b_s + u_s + v_s)
\end{equation}
and redefine the order parameters and the other quantities in the
following way:
\begin{eqnarray}
\label{eq11} &&\varphi_j =b^{1/4}\psi_j =
\phi_je^{i\theta_j}\:,\;\;\; M = b_f^{1/4}{\cal{M}}\:,\\ \nonumber
&& r = \frac{a_s}{\sqrt{b}}\:,\;\;\; t =\frac{a_f}{\sqrt{b_f}}\:,
\;\;\; w = \frac{u_s}{b}\:, \;\;\; v =\frac{v_s}{b}\:, \\
\nonumber &&\gamma= \frac{\gamma_0}{b^{1/2}b_f^{1/4}}\:,\;\;\;
\gamma_1= \frac{\delta}{(bb_f)^{1/2}}.
\end{eqnarray}

Having in mind that the order parameters $\psi$ and
$\mbox{\boldmath$M$}$ are considered uniform and using Eqs.~(10)
and (11), we can write the free energy density
 $f(\psi,M) = F(\psi,M)/V$ in the form
\begin{eqnarray}
\label{eq12} f(\psi,M)& = & r\phi^2 + \frac{1}{2}\phi^4
  + 2\gamma\phi_1\phi_2 M \mbox{sin}(\theta_2-\theta_1)  \\ \nonumber
   && + \gamma_1 \phi^2 M^2
+ tM^2 + \frac{1}{2}M^4\\ \nonumber && -2w
[\phi_1^2\phi_2^2\mbox{sin}^2(\theta_2-\theta_1)
 +\phi_1^2\phi_3^2\mbox{sin}^2(\theta_1-\theta_3) \\ \nonumber && +
 \phi_2^2\phi_3^2\mbox{sin}^2(\theta_2-\theta_3)] \\ \nonumber
&& -v[\phi_1^2\phi_2^2 + \phi_1^2\phi_3^2 + \phi_2^2\phi_3^2].
\end{eqnarray}
Note, that in the above expression the order parameters $\psi$ and
$\mbox{\boldmath$M$}$ are defined per unit volume.

The equilibrium phases are obtained from the equations of state
\begin{equation}
\label{eq13} \frac{\partial f(\mu_0)}{\partial \mu_{\alpha}} =
0\:,
 \end{equation}
 where the series of symbols $\mu$ can be defined as, for example,
 $\mu = \left\{\mu_\alpha\right\}=
 (M, \phi_1,..., \phi_3,$ $ \theta_1,..., \theta_3)$; $\mu_0$ denotes an
equilibrium phase. The stability matrix $\tilde{F}$ of the phases
$\mu_0$
 is defined by
\begin{equation}
\label{eq14}
 \hat{F}(\mu_0)= \left\{F_{\alpha\beta}(\mu_0)\right\} = \frac{\partial^2f(\mu_0)}
{\partial\mu_{\alpha}\partial\mu_{\beta}}\;.
\end{equation}

An alternative treatment can be done in terms of real
($\psi^{\prime}_j$) and imaginary ($\psi^{\prime\prime}_j$) parts
of the complex numbers $\psi_j = \psi_j^{\prime} +
i\psi_j^{\prime\prime}$. The calculation with  moduli $\phi_j$ and
phase angles $\theta_j$ of $\psi_j$  is more simple but
 in cases of strongly degenerate phases  some of
the angles $\theta_j$ remain unspecified. Then  an alternative analysis
with the help of the components $\psi_j^{\prime}$ and
$\psi_j^{\prime\prime}$ should be done.

The thermodynamic stability of the phases that are solutions of
Eqs.~(13) is checked with the help of the matrix~(14). An additional
stability analysis is done by the comparison of  free energies of
phases that  satisfy (13) and render the stability matrix (14) positive
in one and the same domain of parameters $\{r,t,\gamma,\gamma_1,w,v\}$.
This step is important because the complicated form of the free energy
generates a great number of solutions of Eqs.~(13) and we have to sift
out the stable from metastable phases that correspond either to global
or local
 minima of the free energy, respectively ~\cite{Uzunov:1993}.

 Some solutions  of Eqs.~(13) have  a marginal stability, i.e., their
 stability matrix (14) is
  neither positively nor negatively definite.
 This is often a result of the degeneration of
phases with broken continuous symmetry. If the reason for the lack
of a clear positive definiteness of the stability matrix is
precisely the mentioned degeneration of the ground state, one may
reliably conclude that the respective phase is stable. If there is
another reason, the analysis of the matrix (14) will be
insufficient
 to determine the respective stability property. These
cases are quite rare and occur for particular values of the parameters
$\{r,t,\gamma,...\}$.

\section[]{Pure superconductivity}

Let us set $M\equiv 0$ in Eq.~(12) and  briefly summarize the
known results~\cite{Volovik:1985,Blagoeva:1990} for the
 ``pure superconducting case'' when the magnetic order cannot appear and
 magnetic effects do not affect the
 stability of the uniform (Meissner) superconducting phases. The
possible phases can be classified by the structure of the complex
vector order parameter $\psi = (\psi_1,\psi_2,\psi_2)$. We shall
often use the moduli vector $(\phi_1, \phi_2,\phi_3)$ with
magnitude $\phi = (\phi_1^2+\phi_2^2+\phi_3^2)^{1/2}$ and the
phase angles $\theta_j$.

The normal phase (0,0,0) is always a solution of the Eqs.~(13). It
is stable for $r\geq 0$, and corresponds to a free energy $f=0$.
Under certain conditions, six ordered
phases~\cite{Volovik:1985,Blagoeva:1990} occur for $r<0$. Here we
shall not repeat the detailed description of these
phases~\cite{Volovik:1985, Blagoeva:1990} but we shall briefly
mention their structure.

The simplest ordered phase is of type $(\psi_1,0,0)$ with
equivalent domains: $(0,\psi_2,0)$ and $(0,0,\psi_3)$. Multi-
domain phases of more complex structure also occur, but we shall
not always enumerate the possible domains. For example, the
``two-dimensional'' phases
 can be fully represented by domains of type
 $(\psi_1,\psi_2,0)$ but there are also other two types of domains:
 $(\psi_1,0,\psi_3)$ and
$(0,\psi_2,\psi_3)$. As we consider the general case when the
crystal anisotropy is present $(v \neq 0)$, this type of phases
possesses the property
 $|\psi_i| = |\psi_j|$.

The two-dimensional phases are two and have different free
energies. To clarify this point let us consider, for example, the
phase $(\psi_1,\psi_2,0)$. The two complex numbers, $\psi_1$ and
$\psi_2$ can be represented either as two-component real vectors,
or, equivalently, as rotating vectors in the complex plane. One
can easily show that Eq.~(12) yields two phases: a collinear
phase, when $(\theta_2-\theta_1) = \pi k (k = 0,\pm1,...)$, i.e.
when the vectors $\psi_1$ and $\psi_2$ are collinear, and another
(non-collinear) phase when the same vectors are perpendicular to
each other: $(\theta_2-\theta_1) = \pi(k + 1/2)$. Having in mind
that $|\phi_1| = |\phi_2| = \phi/\sqrt{2}$, the domain
$(\psi_1,\psi_2,0)$ of the collinear phase is given by
$(\pm1,1,0)\phi/\sqrt{2}$ whereas the same domain for the
 non-collinear phase will be $(\pm i,1,0)\phi/\sqrt{2}$.
The two domains of these phases have similar representations.

In addition to the mentioned three ordered phases, three other
ordered phases exist. For these phases all three components
$\psi_j$ have nonzero equilibrium values.  Two of them have equal
to one another moduli $\phi_j$, i.e., $\phi_1=\phi_2=\phi_3$. The
third phase is of the type $\phi_1=\phi_2 \neq \phi_3$  and is
unstable so it cannot occur in real systems. The two
three-dimensional phases with equal moduli of the order parameter
components have different phase
 angles and, hence, different structure. The difference between any couple
 of angles $\theta_j$ is given by $\pm \pi/3$ or $\pm 2\pi/3$. The
characteristic vectors of this phase can be of the form
$(e^{i\pi/3}, e^{-i\pi/3},1)\phi/\sqrt{3}$ and $(e^{2i\pi/3},
 e^{-i2\pi/3},1)\phi/\sqrt{3}$. The second stable three dimensional phase
 is ``real'', i.e. the components $\psi_j$ lie on the real axis;
 $(\theta_j-\theta_j) = \pi k$ for any couple of angles $\theta_j$ and the
characteristic vectors are $(\pm 1, \pm 1, 1)\phi/\sqrt{3}$. The
stability properties of these five stable ordered phases were
presented in details in Refs.~\cite{Volovik:1985,Blagoeva:1990}.

When the crystal anisotropy is not present ($v =0$) the picture
changes. The increase of the level of degeneracy of the ordered
states leads to an instability of some phases and to a lack of
some noncollinear phases. Both two- and three-dimensional real
phases, where $(\theta_j -\theta_j) = \pi k$, are no more
constrained by the condition $\phi_i=\phi_j$ but rather have the
freedom of a variation of the moduli $\phi_j$ under the condition
$\phi^2 = -r >0$. The two-dimensional noncollinear phase exists
but has a marginal stability~\cite{Blagoeva:1990}. All other
noncollinear phases even in the presence of a crystal anisotropy
$(v\neq 0)$ either vanish or are unstable; for details, see
Ref.~\cite{Blagoeva:1990}. This discussion demonstrates that the
crystal anisotropy stabilizes the ordering along the main
crystallographic directions, lowers the level of degeneracy of the
ordered state related with the spontaneous breaking of the
continuous symmetry and favors the appearance of noncollinear
phases.

The crystal field effects related to the unconventional
superconducting order were established for the first time in
Ref.~\cite{Volovik:1985}. In our consideration of unconventional
ferromagnetic superconductors in Sec.~4--7 we shall take advantage
of these effects of the crystal anisotropy. In both cases $v=0$
and $v \neq 0$ the matrix (14) indicates an
 instability of three-dimensional phases (all $\phi_j \neq 0)$ with an
 arbitrary ratios $\phi_i/\phi_j$. As already mentioned,
for $v \neq 0$ the phases of type $\phi_1= \phi_2\neq \phi_3$
 are also unstable whereas for $v=0$, even the phase $\phi_1=\phi_2=\phi_3 > 0$
is unstable.

\section[]{M-triggered superconductivity}

We shall consider the Walker-Samokhin model~\cite{Walker:2002}
when only the $M\phi_1\phi_2-$coupling between the order
parameters $\psi$ and $M$ is taken into account ($\gamma > 0$,
$\gamma_1 = 0$) and  the anisotropies $(w=v=0)$ are ignored. The
uniform phases and the phase diagram in this case were
investigated in Refs.~\cite{Shopova1:2003,Shopova2:2003,
Shopova3:2003}.

\small
\begin{table} [hb]
\caption{\label{irreps} Phases and their existence and stability
properties [$\theta = (\theta_2-\theta_1)$, $k = 0, \pm
1,...$].}\smallskip
\begin{small}\centering
\begin{tabular*}{\textwidth}{@{\extracolsep{\fill}}cccc}
\hline  \noalign {\smallskip} N & $\phi_j = M = 0$ & always & $t >
0, r > 0$ \\ \hline\noalign {\smallskip}

FM & $\phi_j = 0$, $M^2 = -t$& $t < 0$& $r>0$, $r > r_e(t)$\\
\hline\noalign {\smallskip}

SC1 & $\phi_1=M=0$, $\phi^2 = -r$ & $r<0$ & unstable  \\
\hline\noalign {\smallskip}

SC2 & $\phi^2= -r$, $\theta = \pi k$, $M = 0$ & $r<0$ & $(t > 0)^*$\\
\hline\noalign {\smallskip}

SC3 & $\phi_1=\phi_2=M=0$, $\phi^2_3 = -r$ & $r<0$ &$r<0$, $t>0$\\
\hline\noalign {\smallskip}

CO1 &$\phi_1= \phi_2=0$, &$r<0$,  &
$r<0$ \\
&$\phi^2_3 = -r$, $M^2=-t$ & $t <0$ & $t < 0$  \\  \hline\noalign
{\smallskip}

CO2 &$\phi_1=0$, $\phi^2 = -r$& $r<0$ & unstable \\
&$\theta=\theta_2=\pi k$, $M^2=-t$ & $t < 0$ & \\ \hline\noalign
{\smallskip}

FS & $2\phi_1^2 = 2\phi_2^2 = \phi^2 = -r + \gamma M$
 & $\gamma M > r$ &  $3M^2>(-t +\gamma^2/2)$ \\
&$\phi_3 = 0$, $\theta= 2\pi(k - 1/4) $ & & $M > 0$\\
& $\gamma r = (\gamma^2-2t)M - 2M^3$ & &  \\ \hline\noalign
{\smallskip}

FS$^{\ast}$ & $2\phi_1^2 = 2\phi_2^2 = \phi^2 = -(r + \gamma M)$,
 & $-\gamma M > r$ &  $3M^2>(-t +\gamma^2/2)$ \\
&$\phi_3 = 0$, $\theta= 2\pi(k + 1/4) $ & & $M < 0$ \\
&$\gamma r = (2t -\gamma^2)M + 2M^3$& & \\ \hline
\end{tabular*}
\end{small}
\end{table}

Here we summarize the main results in order to make a clear
comparison with the results presented in Sec.~5 and Sec.~6. Our
main aim is the description of {\em a trigger effect} which
consists of the appearance of a ``compelled superconductivity''
caused by the presence of ferromagnetic order (here, this is a
standard uniform ferromagnetic order); see also
Refs.~\cite{Shopova1:2003,Shopova2:2003, Shopova3:2003} where this
effect has been already established and briefly discussed. As
mentioned in the Introduction, a similar trigger effect is known
in the physics of improper ferroelectrics. We shall set $\theta_3
\equiv 0$ and use the notation $\theta \equiv \Delta\theta =
(\theta_2 - \theta_1)$.

\subsection[]{Phases}

The possible (stable, metastable and unstable)
 phases are given in Table 1 together with the respective
existence and stability conditions.  The normal or disordered
phase, denoted in Table 1 by $N$, always exists (for all
temperatures $T \geq 0)$ and is stable for $t >0$, $r > 0$. The
superconducting phase denoted in Table 1 by SC1 is unstable. The
same is valid for the phase of coexistence of ferromagnetism and
superconductivity denoted in Table 1 by CO2. The N--phase, the
ferromagnetic phase (FM), the superconducting phases (SC1--3) and
two of the phases of coexistence (CO1--3) are generic phases
because they appear also in the decoupled case $(\gamma\equiv 0)$.
When the $M\phi_1\phi_2$--coupling is not present, the phases
SC1--3 are identical and represented by the order parameter $\phi$
with components $\phi_j$ that participate on  equal footing. The
asterisk attached to the stability condition of the second
superconductivity phase (SC2) indicates that our analysis is
insufficient to determine whether this phase corresponds to a
minimum of the free energy.

It will be shown that the phase SC2, two other purely
superconducting phases and the coexistence phase CO1, have no
chance to become stable for $\gamma \neq 0$. This is so, because
the phase of coexistence of superconductivity and ferromagnetism
(FS in Table 1) that does not occur for $\gamma = 0$ is stable and
has a lower free energy in their domain of stability.  A second
domain $(M < 0)$ of the FS phase is denoted in Table 1 by FS$^*$.
Here we shall describe only the first domain FS. The domain
FS$^{\ast}$ is considered in the same way.

\begin{figure}
\begin{center}
\epsfig{file=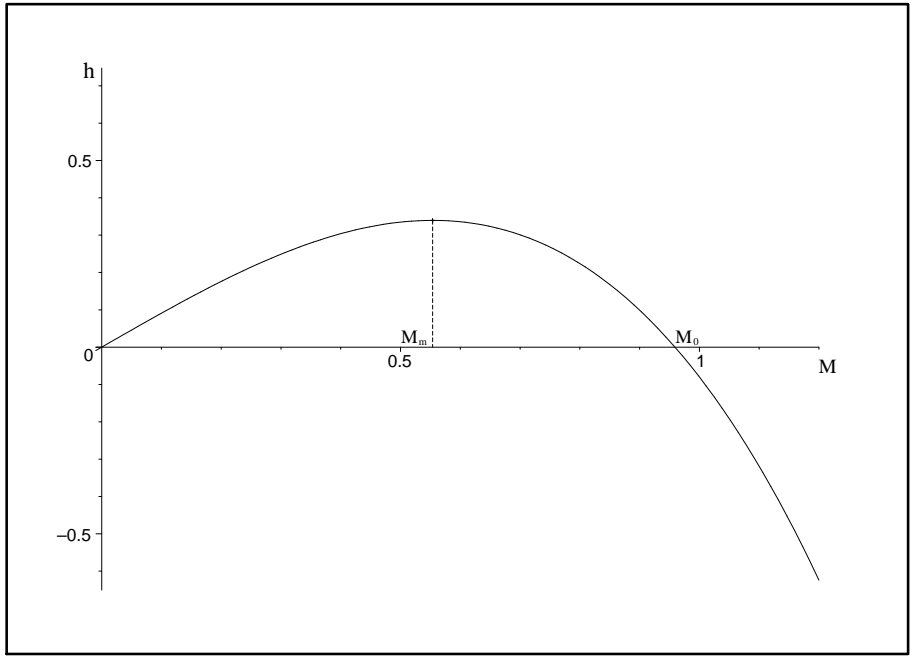,angle=-90, width=12cm}
\end{center}
\caption{$h=\gamma r/2$ as a function of $M$ for $\gamma = 1.2$,
and $t = -0.2$. The parameters $r$, $t$, and $\gamma$ are given by
Eq.~(11).} \label{Uzunovf1.fig}
\end{figure}

\begin{figure}
\begin{center}
\epsfig{file=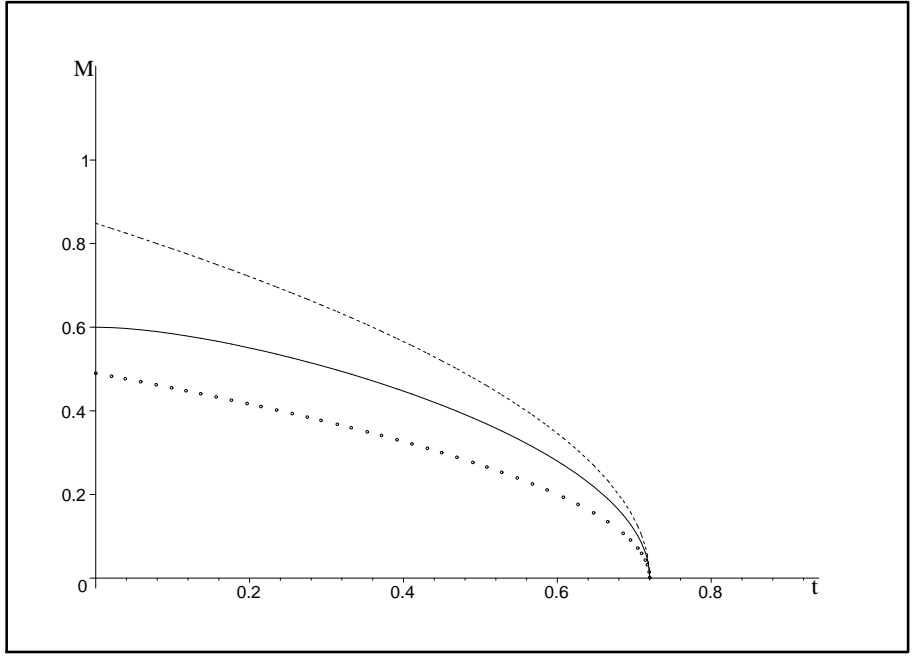,angle=-90, width=12cm}
\end{center}
\caption{The magnetization $M$ versus $t$ for $\gamma = 1.2$: the
dashed line represents $M_0$, the solid line represents $M_{eq}$,
and the dotted line corresponds to $M_m$.} \label{Uzunovf2.fig}
\end{figure}

\begin{figure}
\begin{center}
\epsfig{file=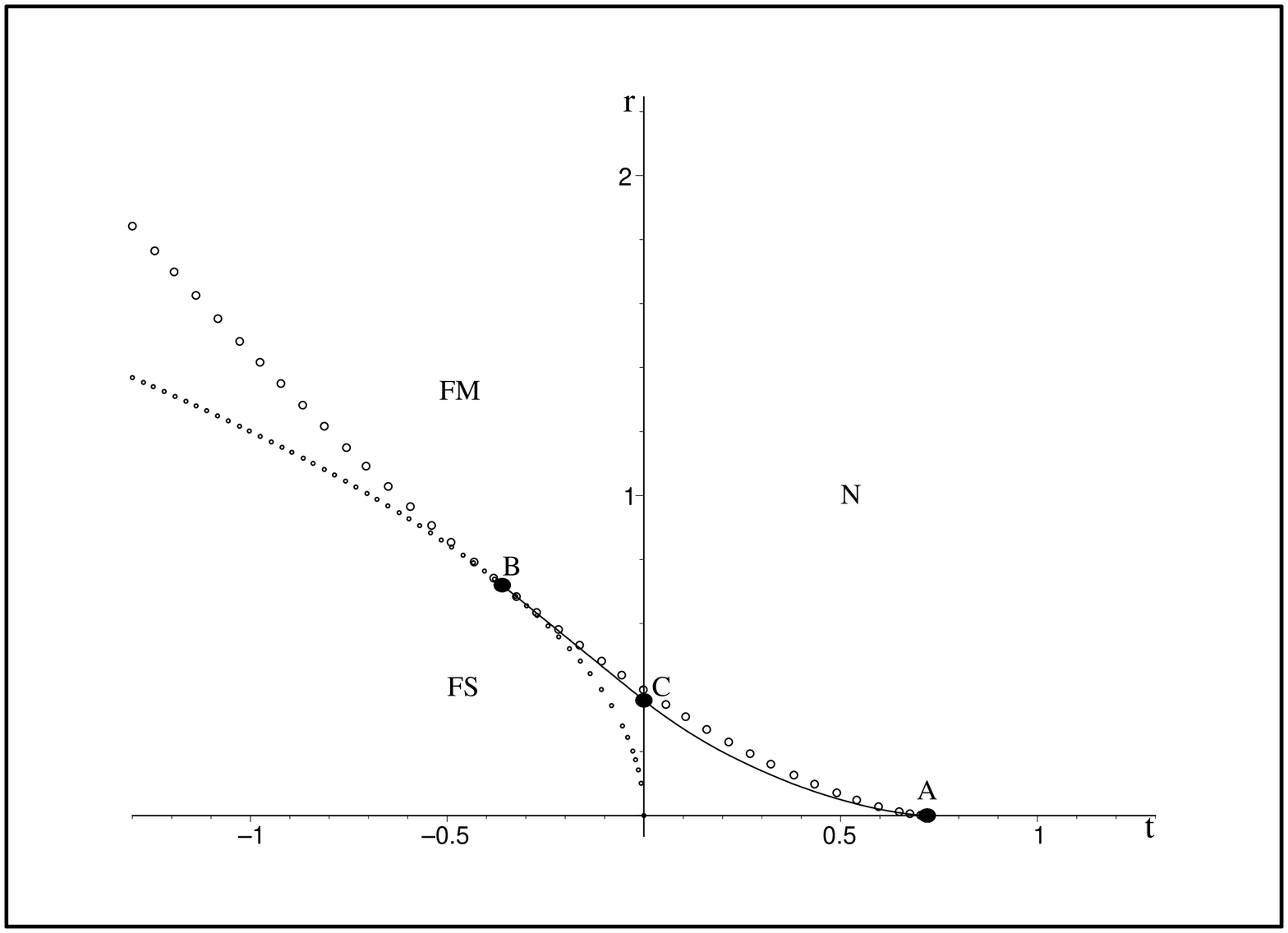,angle=-90, width=12cm}
\end{center}
\caption{The phase diagram in the plane ($t$, $r$) with two
tricritical points (A and B) and a triple point $C$; $\gamma =
1.2$. The parameters $r \sim [T - T_s(P)]$ and $t \sim [T-T_f(P)]$
are defined by Eq.~(11). The domains of existence and stability of
the phases N, FM and FS are shown. The line of circles represents
the function $r_m(t)$ given by Eq.~(17). The dotted line
represents the function $r_e(t)$ given by Eq.~(15). On the left of
point $B$, the same dotted curve corresponds to a FM-FS phase
transition of second order. The equilibrium lines of N-FS and
FM-FS phase transitions of first order are given by the solid
lines $AC$ and $CB$, respectively.} \label{Uzunovf2.fig}
\end{figure}

The cubic equation for magnetization of  FS-phase (see Table~1) is
shown in Fig.~1 for $\gamma = 1.2$ and $t = -0.2$. For any $\gamma
> 0$ and $t$, the stable FS thermodynamic states are given by $r
(M) < r_m = r(M_m)$ for $M > M_m > 0$, where $M_m$ corresponds to
the maximum of the function $r(M)$. The dependence of $M_m(t)$ and
$M_0(t) = (-t + \gamma^2/2)^{1/2} = \sqrt{3}M_m(t)$ on $t$ is
drawn in Fig.~2 for $\gamma = 1.2$.  Functions $r_m(t) =
4M_m^3(t)/\gamma$ for $t < \gamma^2/2$ (depicted by the line of
circles in Fig.~3) and
\begin{equation}\label{eq15}
  r_e(t) = \gamma|t|^{1/2},
\end{equation}
 for $t < 0$  define the borderlines of
stability and existence of FS.

\subsection[]{Phase diagram}

We have outlined the domain in the ($t$, $r$) plane where the FS
phase exists and is a minimum of the free energy. For $r < 0$ the
cubic equation  for $M$ (see Table 1) and the existence and
stability conditions are satisfied for any $M \geq 0$ provided $t
\geq \gamma^2 $. For $ t < \gamma^2$ the condition $M \geq M_0$
have to be fulfilled, here the value
 $M_0 = (-t + \gamma^2/2)^{1/2}$ of $M$ is obtained from $r(M_0) = 0$. Thus
for $r = 0$ the N-phase is stable for
 $t \geq \gamma^2/2$, and FS is stable for $t \leq \gamma^2/2$.
For $r > 0$, the requirement for the stability of FS leads to the
inequalities
\begin{equation}
\label{eq16}
  max\left(\frac{r}{\gamma}, M_m\right) < M < M_0,
\end{equation}
where $M_m = (M_0/\sqrt{3})$ and $M_0$ should be the positive
solution of the cubic equation of state from Table~1; $M_m > 0$
gives a maximum of the function $r(M)$; see also Figs.~1 and 2.

The further analysis  defines the existence and  stability domain
of FS below the line AB denoted by circles (see Fig.~3). In Fig.~3
the curve of circles starts from the point A with coordinates
($\gamma^2/2$, $0$) and touches two other (solid and dotted)
curves at the point B with coordinates ($t_B=-\gamma^2/4$, $r_B=
\gamma^2/2$).  Line of
 circles represents the function
$r(M_m) \equiv r_m(t)$ where
\begin{equation}
\label{eq17}
 r_m(t) = \frac{4}{3\sqrt{3}\gamma} \left (\frac{\gamma^2}{2} -
 t\right)^{3/2}.
\end{equation}
Dotted line represents  $r_e(t)$ defined by Eq.~(15). The
inequality $r < r_m(t)$ is a condition for the stability of FS,
whereas the inequality $r \leq r_e(t)$ for $ (-t) \geq \gamma^2/4$
is a condition for the existence of FS as a solution of the
respective equation of state. This existence condition for FS is
obtained from $\gamma M
> r$ (see Table 1).

In the region on the left of the point B in Fig.~3, the FS phase
satisfies the existence condition $\gamma M > r$ only
 below the dotted line. In the domain confined between the lines of circles
 and the dotted
line on the left of the point B the stability condition for FS is
satisfied but the existence condition is broken. The inequality $r
\geq r_e(t)$ is the stability condition of FM for $ 0 \leq (-t)
\leq \gamma^2/4$. For $(-t) > \gamma^2/4$ the FM phase is stable
for all $r \geq r_e(t)$.

In the region confined by the line of circles AB, the dotted line
for $ 0 < (-t) < \gamma^2/4$, and the $t-$axis, the phases N, FS
and FM have an overlap of stability domains. The same is valid for
FS, the SC phases and CO1 in the third quadrant of the plane ($t$,
$r$). The comparison of the respective free energies for $r < 0$
shows that the stable phase is FS whereas the other phases are
metastable within their domains of stability.

The part of
 the $t$-axis given by $r=0$ and $t > \gamma^2/2$
 is a phase transition line of second order
which describes the N-FS transition. The same transition
 for $0 < t < \gamma^2/2$ is represented by the solid line AC which
is the equilibrium transition line of a first order phase
transition. The equilibrium transition curve is given by the
function
\begin{equation}
\label{eq18}
 r_{eq}(t) =
\frac{1}{4}\left[3\gamma - \left(\gamma^2 + 16t
\right)^{1/2}\right]M_{eq}(t).
\end{equation}
Here
  \begin{equation}
\label{eq19}
 M_{eq}(t) =
\frac{1}{2\sqrt{2}}\left[\gamma^2 - 8t + \gamma\left(\gamma^2 +
 16t \right)^{1/2}\right]^{1/2}
\end{equation}
is the equilibrium jump of the magnetization. The order of the
N-FS transition changes at the tricritical point A.

The domain above the solid line AC and below the line of circles
for $ t > 0$ is the region of a possible
 overheating of FS.
The domain of overcooling of the N-phase is confined by the solid
line AC and the axes ($t > 0$, $r >0$). At the triple point C with
coordinates
 [0, $r_{eq}(0) = \gamma^2/4$]
the phases N, FM, and FS coexist. For $t < 0$ the straight line
\begin{equation}
\label{eq20} r_{eq}^* (t) =  \frac{\gamma^2}{4} + |t|,\;\;\;\;\;\;
t_B < t < 0,
\end{equation}
describes the extension of the equilibrium phase transition line
of the N-FS first order transition to negative values of $t$.
 For $t < t_B$
 the equilibrium phase transition FM-FS is of second order and is
given by the dotted line on the left of the point B which is the
second tricritical point in this phase diagram. Along the first
order transition line
 $r_{eq}^{\ast}(t)$ given by~ Eq.~(\ref{eq20}) the equilibrium value
 of $M$ is $M_{eq} =\gamma/2$,  which
implies an equilibrium order parameter jump at the FM-FS
transition equal to ($\gamma/2 - \sqrt{|t|}$). On the dotted line
of the second order FM-FS
 transition the equilibrium value
of $M$ is equal to that of the FM phase ($M_{eq} = \sqrt{|t|}$).
The FM phase does not exist below $T_s$ and this is a shortcoming
of the model~(\ref{eq12}) with $\gamma_1 = 0$.

The equilibrium  FM-FS and N-FS phase transition lines in Fig.~3
can be expressed by the respective equilibrium phase transition
temperatures $T_{eq}$ defined by the equations $r_e = r(T_{eq})$,
$r_{eq} = r(T_{eq})$, $r^{\ast}_{eq} = r(T_{eq})$, and with the
help of the relation $M_{eq} = M(T_{eq})$. This limits  the
possible variations of parameters of the theory.
 For example, the critical temperature
($T_{eq} \equiv T_c$) of the FM-FS second order transition
 ($\gamma^2/4 < -t$)  is obtained in the form
$T_{c} = (T_s + 4\pi J{\cal{M}}/\alpha_s)$, or, using ${\cal{M}} =
(-a_f/b_f)^{1/2}$,
\begin{equation}
\label{eq21} T_{c} = T_s -\frac{T^{\ast}}{2} + \left[
\left(\frac{T^{\ast}}{2}\right)^2 +
T^{\ast}(T_f-T_s)\right]^{1/2}.
\end{equation}
Here $T_f > T_s$, and $T^{\ast} = (4\pi
J)^2\alpha_f/\alpha_s^2b_f$ is
 a characteristic temperature of the model~(\ref{eq12}) with
 $\gamma_1=w=v=0$. A discussion of Eq.~(21) is given in Sec.~5.3.

The investigation of the conditions for the validity of
Eq.~(\ref{eq21}) leads to the conclusion that the FM-FS continuous
phase transition (at $\gamma^2 < -t)$ will be possible only if the
following condition is satisfied:
\begin{equation}
\label{eq22} T_{f} - T_s > \ = (\varsigma +
\sqrt{\varsigma})T^{\ast},
\end{equation}
where $\varsigma = b_f\alpha_s^2/4b_s\alpha_f^2$.
 Therefore, the second
order FM-FS transition should disappear for a sufficiently large
$\gamma$--coupling. Such a condition does not exist for the first
order transitions FM-FS and N-FS.

 The inclusion of the gradient term (4) in the free
energy~(\ref{eq2}) should lead to a depression of the equilibrium
transition temperature. As the magnetization increases with the
decrease of the temperature, the vortex state should occur at
temperatures which are lower than
 the equilibrium temperature $T_{eq}$ of
the Meissner state. For example, the critical temperature
($\tilde{T}_c$)
 corresponding to the vortex phase
of FS-type has been evaluated~\cite{Walker:2002} to be
 lower than the critical temperature ~(\ref{eq21}): $(T_c - \tilde{T}_c) =
4\pi \mu_B{\cal{M}}/\alpha_s$, where $\mu_B = |e|\hbar/2mc$ is the
Bohr magneton.
 For $J \gg \mu_B$, we have $T_c \approx \tilde{T}_c$.

For $ r > 0$, namely, for temperatures $T > T_s$ the
superconductivity is triggered by the magnetic order through the
$\gamma$-coupling. The superconducting phase for $T > T_s$ is
entirely in the $(t,r)$ domain of the ferromagnetic phase.
Therefore, the uniform supeconducting phase can occur for $T >
T_s$ only through a coexistence with the ferromagnetic order.

The properties of the magnetic susceptibility and the specific
heat near the phase transition lines shown in Fig.~3 have been
investigated in Refs.~\cite{Shopova3:2003, Shopova1:2005} and here
we shall not dwell on these topics. Note that the
results~\cite{Shopova3:2003, Shopova1:2005} for the thermodynamic
quantities should be extended to include the physical effects
considered in the next parts of this review. Such a consideration
requires a numerical analysis.

In the next Sections we shall focus on the temperature range $T >
T_s$ which seems to be of main practical interest. We shall not
dwell on the superconductivity in the fourth quadrant
 $(t >0,r<0)$ of the $(t,r)$ diagram where pure superconducting phases
can occur for systems with $T_s > T_f$,  but this is not the case
for UGe$_2$, URhGe and ZrZn$_2$. Also we shall not discuss the
possible metastable phases in the third quadrant $(t<0,r<0)$ of
the $(t,r)$ diagram.

\subsection[]{Note about a simplified theory}

The analysis in this Section can be done following an approximate
scheme known from the theory of improper ferroelectrics; see,
e.g., Ref.~\cite{Cowley:1980}.  In this approximation the order
parameter $M$  is considered small enough which makes possible to
ignore $M^4$-term in the free energy. Then one easily obtains from
the data for FS presented in Table~1 or by a direct calculation of
the respective reduced free energy that the order parameters
$\phi$ and $M$ of FS--phase are described by the simple equalities
$r = (\gamma M -\phi^2)$ and $M = (\gamma/2t)\phi^2$.  For
ferroelectrics working with oversimplified free energy
 gives a substantial departure of theory from experiment~\cite{Cowley:1980}.
 The same approximation has been recently applied to ferromagnetic Bose-Einstein
condensates~\cite{Gu:2003, Gu:2005}.

For ferromagnetic superconductors the domain of reliability of
this approximation could be the close vicinity of the
ferromagnetic phase transition, i.e., for temperatures near the
critical temperature $T_f$. This discussion can be worthwhile if
only the primary order parameter also exists in the same narrow
temperature domain ($\phi > 0$). Therefore, the application of the
simplified scheme can be useful in systems, where $T_s \ge T_f$.

 For $T_s<T_f$,
the analysis can be simplified if we suppose  a relatively small
value of the modulus $\phi$ of the superconducting order
parameter. This approximation should be valid in some narrow
temperature domain near the line of second order phase transition
from FM to FS.

\section[]{Effect of symmetry conserving coupling}

Here we shall include in our consideration   both linear and
quadratic couplings of magnetization to the superconducting order
parameter which means that both parameters $\gamma$ and $\gamma_1$
in free energy~(12) are different from zero. In this way we shall
investigate the effect of the symmetry conserving $\gamma_1$-term
in the free energy on the thermodynamics of the system. When
$\gamma$ is equal to zero but $\gamma_1 \ne 0$ the analysis is
easy and the results are known from the theory of bicritical and
tetracritical points~\cite{Uzunov:1993, Toledano:1987, Liu:1973,
Imry:1975}. For the problem of coexistence of conventional
superconductivity and ferromagnetic order the analysis $(\gamma =
0, \gamma_1 \neq 0)$ was made in Ref.~\cite{Vonsovsky:1982}.

 At this
stage we shall not take into account any anisotropy effects
 because we do not want to obscure the influence of quadratic
interaction by considering too many parameters. For
$\gamma,\gamma_1 \ne
 0$ and $w=0$, $v=0$ the results again  can be presented in an analytical form,
 only a small
part of phase diagram should be calculated numerically.

\subsection[]{Phases}

The calculations show that for temperatures $T > T_s$, i.e., for
$r > 0$, we have again three stable phases. Two of them are quite
simple: the normal ($N$-) phase with existence and stability
domains shown in Table~1, and the FM phase with the existence
condition $ t<0$ as shown in Table~1, and a stability domain
defined by the inequality $r_e^{(1)} \le r$. Here
\begin{equation}
\label{eq23} r_e^{(1)}=\gamma_1t + \gamma\sqrt{-t},
\end{equation}
and one can compare it with the respective expression~(15) for
$\gamma_1=0$. In this paragraph we shall retain the same notations
as in Sec.~4, but with a superscript $(1)$ in order to distinguish
them from the case $\gamma_1=0$ The third stable phase for $r>0$
is a more complex variant of the mixed phase FS and its domain
FS$^*$, discussed in Sec.~4. The symmetry of the FS phase
coincides with that found in ~\cite{Walker:2002}.

We have to mention that for $r<0$ there are five pure
superconducting ($M =0$, $\phi
> 0$) phases. Two of them, $(\phi_1 > 0, \phi_2 =
\phi_3 =0)$ and $(\phi_1 =0, \phi_2>0, \phi_3>0)$ are unstable.
Two other phases, $(\phi_1>0, \phi_2>0, \phi_3 =0, \theta_2 =
\theta_1 + \pi k)$ and $(\phi_1>0,\phi_2>0, \phi_3>0, \theta_2 =
\theta_1 + \pi k, \theta_3$ -- arbitrary; $k=0,\pm1,...)$ show a
marginal stability for $ t > \gamma_1 r$.

Only one of the five pure superconducting phases,  the phase SC3,
given in Table~1, is stable. In case of $\gamma_1 \neq 0$ the
values of $\phi_j$ and the existence domain of SC3 are the same as
shown in Table ~1 for $\gamma_1 =0$ but the stability domain is
different and is given by $t > \gamma_1 r$. When the anisotropy
effects are taken into account the phases exhibiting marginal
stability within the present approximation may become stable.
Besides, three other mixed phases $(M \neq 0, \phi
>0)$ exist for $r < 0$ but one of them is metastable (for
$\gamma_1^2 >1, t < \gamma_1 r$, and $r < \gamma_1 t$) and the
other two are absolutely unstable.

Here the thermodynamic behavior for $r < 0$ is much more abundant
in phases than for improper ferroelectrics with two component
primary order parameter ~\cite{Toledano:1987}. However, at this
stage of experimental needs about the properties of unconventional
ferromagnetic superconductors the investigation of the phases for
temperatures $T < T_s$ is not of primary interest and for this
reason we shall focus our attention on the temperature domain $r >
0$.

The FS phase for $\gamma_1 \ne 0$ is described by the following
equations:
\begin{equation}
\label{eq24} \phi_1 = \phi_2=\frac{\phi}{\sqrt{2}}\:, \;\;\;
\phi_3 = 0,
\end{equation}
\begin{equation}
\label{eq25} \phi^2= (\pm \gamma M-r-\gamma_1 M^2),
\end{equation}
\begin{equation}
\label{eq26} (1-\gamma_1^2)M^3\pm \frac{3}{2} \gamma \gamma_1 M^2
+\left(t-\frac{\gamma^2}{2}-\gamma_1 r\right)M \pm \frac{\gamma
r}{2}=0,
\end{equation}
and
\begin{equation}
\label{eq27} (\theta_2 - \theta_1) = \mp \frac{\pi}{2} + 2\pi k,
\end{equation}
($k = 0, \pm 1,...$). The upper sign in Eqs.~(24)~-~(27)
corresponds to the FS domain  where $\mbox{sin}(\theta_2-\theta_1)
= -1$ and the lower sign corresponds to the FS$^{*}$ domain with
$\mbox{sin}(\theta_2-\theta_1) = 1$. This is a generalization of
the two-domain  FS phase discussed in Sec.~3. The analysis of the
stability matrix (14) for these phase
 domains shows that FS is stable for $M > 0$ and FS$^{*}$ is stable for
$M<0$, just like our result in Sec.~4. As these domains belong to
the same phase, namely, have the same free energy and are
thermodynamically equivalent, we shall consider one of them, for
example, FS.

\subsection[]{Phase stability and phase diagram}

In order to outline the ($t,r$) phase diagram  we shall use the
 information given above for the other two phases which have their own
 domains of stability in the $(t,r)$ plane: N and FM. The
 FS stability conditions when $\gamma_1 \ne 0$ become

\begin{equation}
\label{eq28} 2 \gamma M -r -\gamma_1 M^2\ \ge 0,
\end{equation}
\begin{equation}
\label{eq29} \gamma M \ge 0,
\end{equation}
\begin{equation}
\label{eq30} 3(1-\gamma_1^2)M^2+3\gamma \gamma_1 M+t-\gamma_1
r-\gamma^2/2 \ge 0.
\end{equation}

and we prefer to  treat Eqs.~(28)~-~(30)~together with the
existence condition $\phi^2 \ge 0$, with $\phi$ given by Eq.~(25),
with the help of the picture shown in Fig.~4.

The most direct approach to analyze the existence and stability of
FS phase is to
 express $r$ as of function of $(M,t)$ from the equation of state (26),

\begin{eqnarray}
\label{31} \lefteqn{r^{(1)}_{eq}(t)=
}\\\nonumber&&\frac{M_{eq}}{(\gamma_1
M_{eq}-\gamma/2)}\left[(1-\gamma_1^2)M_{eq}^2+ \frac{3}{2} \gamma
\gamma_1 M_{eq} +(t-\frac{\gamma^2}{2})\right],
\end{eqnarray}

 and to substitute the above expression in the existence and
 stability conditions of FS-phase. It is obvious
 that there is a special value of $M$
\begin{equation}\label{32}
M_{S1}=\frac{\gamma}{2 \gamma_1}
\end{equation}
that is
 a solution of Eq.~(26) for any
 value of $r$ and
\begin{equation}\label{33}
  t_{S1}=-\frac{\gamma^2}{4\gamma_1^2},
\end{equation}
for which this procedure cannot be applied and should be
considered separately.  Note, that $M_{S1}$ is given by the
respective horizontal dashed line in Fig.~4.
 The analysis shows that in the interval $t_B^{(1)} < t<\gamma^2/2$ the phase
transition   is again of first order; here
\begin{equation}\label{34}
t_B^{(1)} =-\frac{\gamma^2}{4(1+\gamma_1)^2}.
\end{equation}
To find the equilibrium magnetization of first order phase
transition, depicted by the
  thick line $ACB$
 in Fig.~4 we need  the expression for equilibrium
free energy of FS-phase. It is obtained  from Eq.~(12) by setting
$(w=0,v=0)$ and substituting $r,\;\phi_i$ as given by Eqs.~(24),
(25) and (31). The result is
\begin{eqnarray}
\label{eq35} f^{(1)}_{FS} &=&
-\frac{M^2}{2(M\gamma_1-\gamma/2)^2}\times
 \{(1-\gamma_1^2)M^4 + \gamma\gamma_1 M^3  \\ \nonumber \newline
 &&
+ 2[t(1-\gamma_1^2)- \frac{\gamma^2}{8}] M^2
 - 2\gamma\gamma_1t M + t(t-\frac{\gamma^2}{2})\},
 \end{eqnarray}
where $M \equiv M_{eq}$.

For the phase transition from N to FS phase ($0<t<\gamma^2/2$),
$M_{eq}$  is found by setting the FS  free energy  from the above
expression equal to zero, as we have  by convention that  the free
energy of the normal phase is  zero. The value of $M^{(1)}_{eq}$
for positive $t$ is obtained numerically and is illustrated by
thick black curve $AC$ in Fig.~4. When
 $t^{(1)}_B\le t <0$ the transition is between FM and FS phases
 and we obtain $M^{(1)}_{eq}$ from the equation
 $f_{FS}=f_{FM}=(-t^2/2)$, where $f_{FM}$ is the free energy of FM
 phase. The equilibrium magnetization in the above $t$-interval is
 given by the formula
 \begin{equation}
\label{eq36} M_{eq}^{(1) \ast}=\frac{\gamma}{2(1+\gamma_1)},
\end{equation}
and  is drawn by thick line $CB$ in Fig. 4. The existence and stability
 analysis shows
that for $r>0$ the equilibrium magnetization of the first order phase
 transition  should satisfy
the condition $ M^{(1)}_m < M^{(1)}_{eq}<M^{(1)}_0$.

By $M^{(1)}_0$ we denote  the positive solution of
$r^{(1)}(M_{eq})=0$ and its $t$-dependence is drawn in Fig.~4 by
the curve with circles. $M^{(1)}_m$ is the smaller positive root
of stability condition~(30) and also gives the maximum of the
function $r^{(1)}_{eq}(M)$; see Eq.~(31). The  function
$M^{(1)}_m$ is depicted by the dotted curve  $AB$ in Fig.~4. When
$t_{S1}< t<t^{(1)}_B$ the existence and stability conditions are
fulfilled if $\sqrt{-t}<M<M_{S1}$, where $\sqrt{-t}$ is the
magnetization of ferromagnetic phase and is drawn by a thin black
line on the left of point B in Fig.~(4). Here we have two
possibilities: $r>0$ for $\sqrt{-t}<M <M^{(1)}_0$ and $r < 0$ for
$M^{(1)}_0<M<M_{S1}$. To the left of $t_{S1}$ and $t>t_{S2}$,
where
\begin{equation}\label{eq37}
  t_{S2}= - \left(\frac{\gamma}{\gamma_1}\right)^2.
\end{equation}
the FS phase is stable and exists
 for $M_{S1}<M<\sqrt{-t}$. Here $r$ will be positive when
$M^{(1)}_0<M<\sqrt{-t}$ and  $r<0$ for $M^{(1)}_0>M>M_{S1}$. When
$t<t_{S2}$, $M<\sqrt{-t}$ and $r$ is always negative.

\begin{figure}
\begin{center}
\epsfig{file=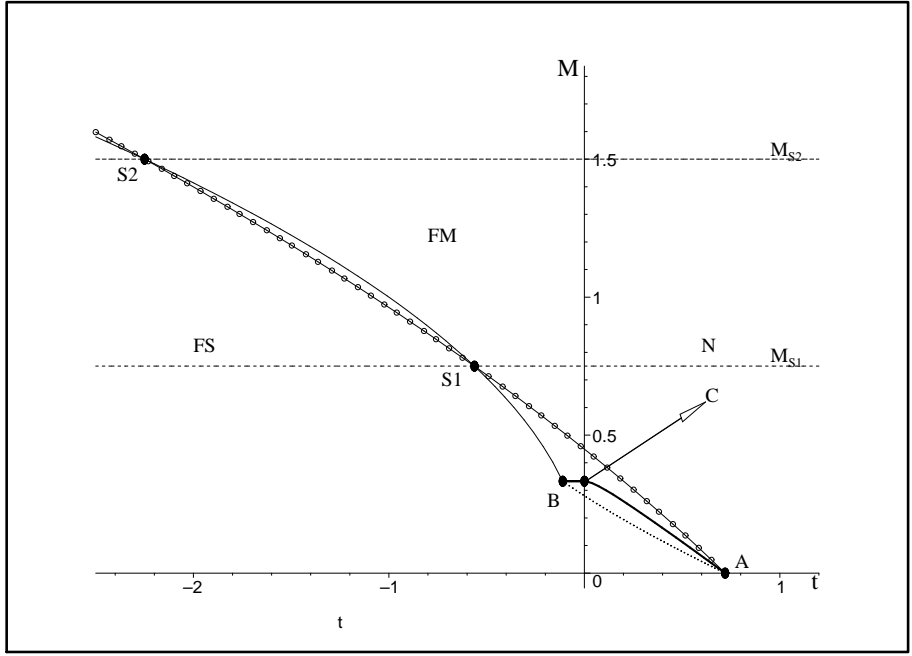,angle=-90, width=12cm}
\end{center}
\caption{The dependence $M(t)$ as an illustration of stability
analysis for $\gamma=1.2$, $\gamma_1=0.8$ and $w=0$. The
parameters of the theory ($r,t,\gamma$, $\gamma_1$, $w,\dots$) are
defined by Eq.~(11). The horizontal dashed lines represent the
quantities $M_{\scriptsize S1}$ given by Eq.~(32) and
$M_{\scriptsize S2} = 2M_{\scriptsize S1}$. The line of circles
$AS_1S_2$ describes the positive solution of Eq.~(31). The thick
line $AC$ gives the equilibrium magnetization for $t>0$. The thick
line $BC$ represents the equilibrium magnetization for $t<0$ as
given by Eq.~(35). The dotted curve is the smaller positive
solution of the stability condition (30). The thin solid line
$BS_1S_2$ is the magnetization $M=\sqrt{-t}$. The arrow indicates
the triple point $C$. $A$ and $B$ are tricritical points of phase
transition. The point $S_1$ corresponds to the maximum of the
curve (23) for $t < 0$, and the point $S_2$ corresponds to
$r_e^{(1)}(t) = 0$ in Eq.~(23).} \label{Uzunovf2.fig}
\end{figure}

\begin{figure}
\begin{center}
\epsfig{file=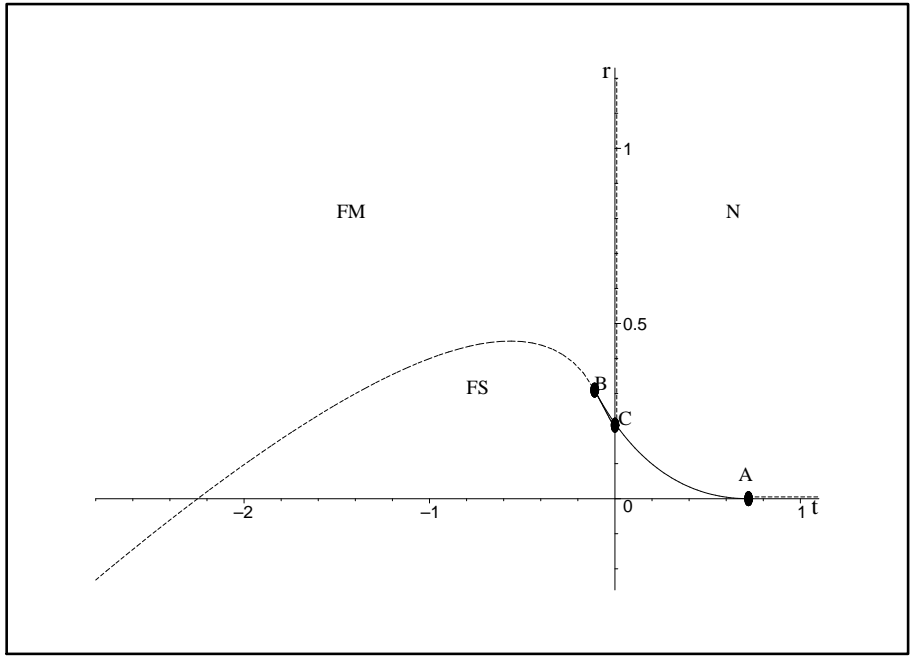,angle=-90, width=12cm}
\end{center}
\caption{The phase diagram in the $(t,r)$ plane for
$\gamma=1.2,\;\gamma_1=0.8$ and $w=0$. The parameters of the
theory ($r$, $t$, $\gamma$, $\gamma_1$, $w,\dots$) are defined by
Eq.~(11). The domains of stability of the phases N, FM and FS are
indicated. $A$ and $B$ are tricritical points of phase transitions
separating the dashed lines (on the left of point $B$ and on the
right of point $A$) of second order phase transitions from the
solid line $ABC$ of first order phase transitions. The FS phase is
stable in the whole domain of the ($t$, $r$) below the solid and
dashed lines. The vertical dashed line coinciding with the
$r$-axis above the triple point $C$ indicates the N-FM phase
transition of second order.} \label{Uzunovf2.fig}
\end{figure}

On the basis of  the existence and stability analysis
 we draw in Fig.~5 the $(t,r)$-phase diagram  for concrete
values of $\gamma$ and $\gamma_1$. As we have mentioned above the
order of phase transitions is the same as for $\gamma_1=0$, see
Fig.~3, Sec.~4.
 The phase transition between the normal and FS phases is of
first order and goes along the equilibrium line $AC$ in the
interval ($t_A=\gamma^2/2 $ and $t_C=0$). The function
$r^{(1)}_{eq}(t)$ is given by Eq. (31) with $M^{(1)}_{eq}$ from
Fig. 4.

N, FM, and FS phases coexist at the triple point $C$ with
coordinates $t=0$, and $r_C^{(1)}=\gamma^2/4(\gamma_1+1)$. On the
left of $C$ for $t^{(1)}_B<t<0$ the phase transition line of first
order
 $r^{(1)\ast}_{eq}(t)$ is found by substituting in Eq.~(31) the
 respective equilibrium magnetization, given by Eq. (35). In result we
 obtain
\begin{equation}
\label{eq38}
  r^{(1)\ast}_{eq}(t) = \frac{\gamma^2}{4(1+\gamma_1)}-t.
\end{equation}
This function is illustrated by the line $BC$ in Fig.~5 that
terminates at the tricritical point $B$ with coordinates $
t^{(1)}_B $ from Eq.~(34), and
\begin{equation}\label{eq39}
  r^{(1)}_B=\frac{
\gamma^2(2+\gamma_1)}{4(1+\gamma_1)^2}.
\end{equation}
To the left of the tricritical point $B$ the second order phase
transition curve is given by the relation~(23). Here
 the magnetization is $M=\sqrt{-t}$ and the
superconducting order parameter is equal to zero
 ($\phi=0$). This line
intersects t-axis at $t_{S2}$ and is well defined also for $r<0$.
The function $r^{(1)}_{e}(t)$ has a maximum at the point $(t_{S1},
\gamma^2/4\gamma_1)$; here $M=M_{S1}$. When this point is
approached the second derivative of the free energy with respect
to $M$ tends to infinity. The result for the curves
$r^{(1)}_{eq}(t)$ of equilibrium phase transitions (N-FS and
FM-FS) can be used to define the respective equilibrium phase
transition temperatures $T_{FS}$.

We shall not discuss the region, $t>0$, $r<0$, because we have
supposed from the very beginning that the transition temperature
for the ferromagnetic ordering T$_f$ is higher then the
superconducting transition temperature T$_s$, as is for the known
unconventional ferromagnetic superconductors. But this case may
become of substantial interest when, as one may expect, materials
with T$_f < $T$_s$ may be discovered experimentally.

\subsection[]{Discussion}

The shape of the equilibrium phase transition lines corresponding
to the phase transitions N-SC, N-FS, and FM-FS is similar to that
of the more simple case $\gamma_1 = 0$ and we shall not dwell on
the variation of the size of the phase domains with the variations
of the parameter $\gamma_1$ from zero to values constrained by the
condition $\gamma_1^2 <1$. We shall draw the attention to the
important qualitative difference between the equilibrium phase
transition lines shown in Figs.~3 and 5. The second order phase
transition line $r_e(t)$, shown by the dotted line on the left of
point $B$ in Fig.~3, tends to large positive values of $r$ for
large negative values of $t$ and remains in the second quadrant
($t<0, r>0)$ of the plane ($t,r$) while the respective second
order phase transition line $r^{(1)}_{e}(t)$ in Fig.~5 crosses the
$t$-axis at the point $t_{S2}$ and is located in the third
quadrant ($t<0,r<0$) for all possible values $t < t_{S2}$. This
means that the ground state (at 0 K) of systems with $\gamma_1 =0$
will be always the FS phase while two types of ground states, FM
and FS, can exist for systems with $0< \gamma_1^2 < 1$. The latter
seems more realistic when we compare theory and experiment,
especially, in ferromagnetic compounds like UGe$_2$, URhGe, and
ZrZn$_2$ where the presence of FM phase is observed at very low
temperatures and relatively low pressure $P$.

The final aim of the phase diagram investigation is the outline of
the ($T,P$) diagram. Important conclusions about the shape of the
$(T,P)$ diagram can be made from the form of the $(t,r)$ diagram
without an additional information about the values of the relevant
material parameters $(a_s$, $a_f,...$) and their dependence on the
pressure $P$. One should know also the characteristic temperature
$T_s$, which has a lower value than the experimentally
observed~\cite{Saxena:2000, Huxley:2001, Tateiwa:2001,
Pfleiderer:2001, Aoki:2001} phase transition temperature $(T_{FS}
\sim 1 K)$ to the coexistence FS--phase. A supposition about the
dependence of the parameters $a_s$ and $a_f$ on the pressure $P$
was made in Ref.~\cite{Walker:2002}. Our results for $T_f \gg T_s$
show that the phase transition temperature $T_{FS}$ varies with
the variation of the system parameters $(\alpha_s, \alpha_f,...)$
from values which are  higher than the characteristic temperature
$T_s$ down to zero temperature. This is seen from Fig.~5.

In systems where a pure superconducting phase is not observed for
temperatures $T \sim T_f$ or $T\sim T_{\scriptsize FS}$, we can
set $T_s \sim 0$ in Eq. (21). Neglecting $T_s$ in Eq.~(21) and
assuming that $(T^{\ast}/T_f) \ll 1$  we obtain that $T_c \equiv
T_{\scriptsize FS} \sim (T^{\ast}T_f)^{1/2}$. Note that the first
$(T^{\ast}/T_f)^{1/2}$-correction to this result has a negative
sign which means that a suitable dependence of the characteristic
temperature $T^{\ast}$ on the pressure P may be used in attempts
to describe the experimental shape of the FM-FS phase transition
line in the $(T,P)$ diagrams of UGe$_2$ and ZrZn$_2$; see, for
example, Fig. 2 in Ref.~\cite{Saxena:2000}, Fig. 3 in
Ref.~\cite{Huxley:2001}, Fig. 4 in Ref.~\cite{Pfleiderer:2001}.
The experimental phase diagrams indicate that $T_f(P)$ is a smooth
monotonically decreasing function of the pressure $P$ and $T_f(P)$
tends to zero when the pressure $P$ exceeds some critical value
$P_c \sim 1$ GPa. Postulating the respective experimental shape of
the function $T_f(P)$ one may try to give a theoretical prediction
for the shape of the curve $T_{\scriptsize FS}$.

The lack of experimental data about
important parameters of the theory forces us to make some
suppositions about the behavior of the function $T^{\ast}(P)$. The
phase transition temperature $T_{\scriptsize FS}$ will
qualitatively follow the shape of $T_f(P)$ provided the dependence
$T^{\ast}(P)$ is very smooth. This is in accord with the
experimental shapes of these curves near the critical pressure
$P_c$ where both $T_f$ and $T_{\scriptsize FS}$ are very small.
The substantial difference between $T_f$ and $T_{\scriptsize FS}$
at lower pressure ($P < P_c$) can be explained with the negative
sign of the correction term to the leading dependence
$T_{\scriptsize FS}(P) \sim [T^{\ast}(P)T_f(P)]^{1/2}$ mentioned
above and a convenient supposition for the form of the function
$T^{\ast}(P)$.

Eq.~(21) presents a rather simplified theoretical result for $T_C
\equiv T_{\scriptsize FS}$ because the effect of $M^2|\psi|^2$
coupling is not taken into account. But following the same ideas,
used in our discussion of Eq.~(21), a more reliable theoretical
prediction of the shape of FM-FS phase transition line can be
given on the basis of Eq.~(23). With the help of the
experimentally found shape of $T_f(P)$ and the definition of the
parameters $r$ and $t$ by Eq.~(11) we can substitute
$T=T_{\scriptsize FS}(P)$ in Eq.~(23). In doing this we have
applied the following approximations, namely, that $T_s \sim 0$
for any pressure $P$, $T_{\scriptsize FS}(P_c) \sim T_f(P_c) \sim
0$ and for substantially lower pressure ($P < P_c$),  $T_f (P) \gg
T_{\scriptsize FS}(P)$. Then near the critical pressure $P_c$, we
easily obtain the transition temperature $T_{\scriptsize FS} \sim
0$, as should be. For substantially lower values of the pressure
there exists an experimental requirement $(T_{\scriptsize FS} -
T_s) \ll (T_f - T_{\scriptsize FS})$. Using the latter we
establish the approximate formula
\begin{equation}
\label{eq40}
(T_f - T_{\scriptsize FS}) =
\gamma^2b_f^{1/2}/\gamma_1^2\alpha_f.
\end{equation}
The same formula for $(T_f
-T_{\scriptsize FS})$ can be obtained from the parameter
$t_{\scriptsize S2}(T_{\scriptsize FS})$ given by Eq.~(37). The
pressure dependence of the parameters included in this formula
defines two qualitatively different types of behavior of
$T_{\scriptsize FS}(P)$ at relatively low pressures ($P \ll P_c$):
(a) $T_{\scriptsize FS}(P) \sim 0$ below some (second) critical
value of the pressure ($P_c^{\prime} < P_c$), and (b) finite
$T_{\scriptsize FS}(P)$ up to $P \sim 0$. Therefore, we can
estimate the value of the pressure $P_c^{\prime} < P_c$ in
UGe$_2$, where $T_{\scriptsize FS}(P_c^{\prime}) \sim 0$. It can
be obtained from the equation
\begin{equation}
\label{eq41}
T_f(P_c^{\prime}) =
(\gamma^2b_f^{1/2}/\gamma_1^2\alpha_f)
\end{equation}
provided the pressure
dependence of the respective material parameters is known. So, the
above consideration is consistent with the theoretical prediction
that the dashed line in Fig.~5 crosses the axis $r=0$ and for this
reason we have the opportunity to describe two ordered phases at
low temperatures and broad variations of the pressure. Our theory
allows also a description of the shape of the transition line
$T_{\scriptsize FS}(P)$ in ZrZn$_2$ and URhGe, where the
transition temperature $T_{\scriptsize FS}$ is finite at ambient
pressure. To avoid a misunderstanding, let us note that the
diagram in Fig.~5 is quite general and the domain containing the
point $r=0$ of the phase transition line for negative $t$ may not
be permitted in some ferromagnetic compounds.

Up to now we have discussed experimental curves of second order
phase transitions. Our analysis gives the opportunity to describe
also first order phase transition lines. Our investigation of the
free energy (12) leads to the prediction of triple ($C$) and
tricritical points ($A$ and $B$); see Figs. 3 and 5. We shall not
consider the possible application of these results to the phase
diagrams of real substances, for which first order phase
transitions and multicritical phenomena occur; see, e.g.,
Refs.~\cite{Kotegawa:2004, Huxley:2003}, where first order phase
transitions and tricritical points have been observed. The
explanation of the phase transition lines in
Refs.~\cite{Kotegawa:2004, Huxley:2003} requires further
theoretical studies that can be done on the basis of a convenient
extension of the free energy (12). For example, the investigation
of vortex phases in Ref.~\cite{Huxley:2003} requires to take into
account the gradient terms (4). Another generalization should be
done in order to explain the observation of two FM
phases~\cite{Kotegawa:2004, Huxley:2003}. Note, that the
experimentalists are not completely certain whether the FS phase
is a uniform or a vortex phase, and this is a crucial point for
the further investigations. But we find quite encouraging that our
studies naturally lead to the prediction of the same variety of
phase transition lines and multicritial points that has been
observed in recent experiments~\cite{Kotegawa:2004, Huxley:2003}.

\section[]{Anisotropy}

Our analysis demonstrates that when the anisotropy of  Cooper
pairs is taken into account, there will be no drastic changes in
the shape the phase diagram for $r>0$ and the order of the
respective phase transitions. Of course, there will be some
changes in the size of the phase domains and the formulae for the
thermodynamic quantities. It is readily seen from Figs.~6 and 7
that the temperature domain of first order phase transitions and
the temperature domain of stability of FS above $T_s$ essentially
vary with the variations of the anisotropy parameter $w$. The
parameter $w$ will also insert changes in the values of the
thermodynamic quantities like the magnetic susceptibility and the
entropy and specific heat jumps at the phase transition points.

Besides, and this seems to be the main anisotropy effect, the $w$-
and $v$-terms in the free energy lead to a stabilization of the
order along the main crystal directions which, in other words,
means that the degeneration of the possible ground states (FM, SC,
and FS) is considerably reduced. This means also a smaller number
of marginally stable states.

\begin{figure}
\begin{center}
\epsfig{file=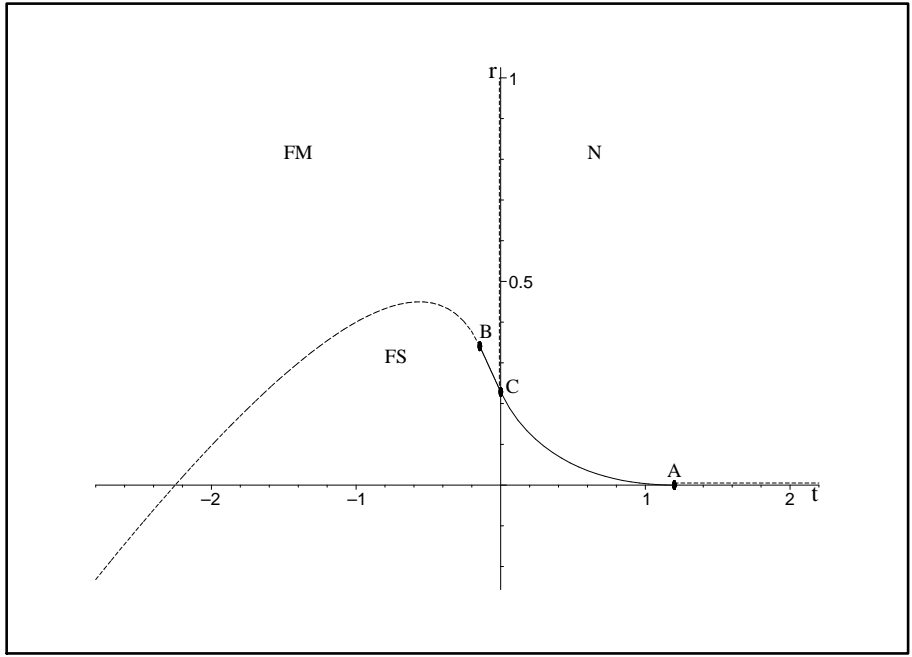,angle=-90, width=12cm}
\end{center}
\caption{Phase diagram in the ($t$, $r$) plane for $\gamma = 1.2,
$ $\gamma_1 = 0.8$, and $w = 0.4$. The meaning of lines and points
is the same as given in Fig.~5.} \label{Uzunovf6.fig}
\end{figure}

\begin{figure}
\begin{center}
\epsfig{file=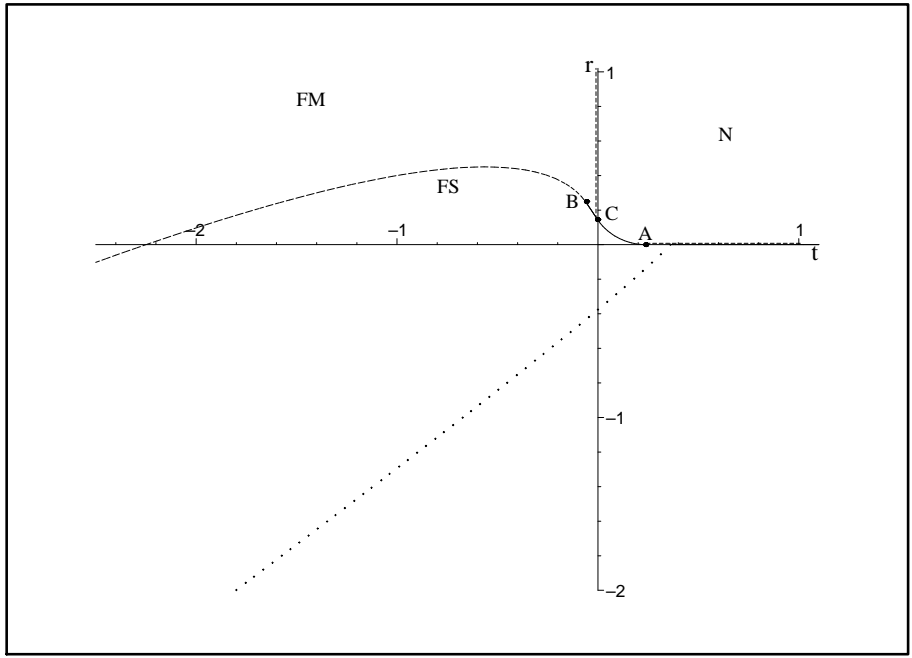,angle=-90, width=12cm}
\end{center}
\caption{Phase diagram in the ($t$, $r$) plane for $\gamma = 1.2,
$ $\gamma_1 = 0.8$, and $w = -2$. The straight dotted line for $r
< 0$ indicates an instability of the FS phase. The meaning of
other lines and notations is the same as given in Fig.~5.}
\label{Uzunovf7.fig}
\end{figure}

The dimensionless anisotropy parameter $w=u_s/(b_s + u_s)$ can be
either positive or negative depending on the sign of $ u_s$.
Obviously when $ u_s > 0$, the parameter $w$ will be positive too
and will be in the interval $0<$ w$<1$ to ensure the positiveness
of parameter $b$ from Eq.~(10).  When $w<0$, the latter condition
is obeyed if the original parameters of free energy (3) satisfy
the inequality $-b_s<u_s<0$.

We should mention here that a new phase of coexistence of
superconductivity and ferromagnetism occurs as a solution of
Eqs.~(13). It is defined in the following way:
\begin{equation}
\label{eq42}
\phi^2_1+\phi^2_2  =
\frac{1}{1-\gamma_1^2}\left[\gamma_1\left(t+\frac{\gamma^2}{2w}\right)
-r\right],
\end{equation}
\begin{equation}
\label{eq43}
 M^2=\frac{1}{1-\gamma_1^2}\left[\gamma_1r-\left(t+\frac{\gamma^2}{2w}\right)
\right],
\end{equation}
and
\begin{equation}
\label{eq44}
 2w\sin(\theta_2-\theta_1)=\gamma M,\;\;\;\;
\cos(\theta_2-\theta_1)\ne 0.
\end{equation}
 In the present approximation the phase~(42) - (44) is unstable, but this may be
changed when the crystal anisotropy is taken into account.

We shall write the equations for order parameters $M$ and $\phi_j$ of FS phase
in order to illustrate the changes when $w\ne 0$
 \begin{equation}
\label{eq45} \phi^2=\frac{\pm \gamma M-r-\gamma_1 M^2}{(1-w)} \ge
0,
\end{equation}
and
\begin{eqnarray}
\label{eq46} \lefteqn{(1- w - \gamma_1^2)M^3}\\\nonumber&&\pm
\frac{3}{2} \gamma \gamma_1 M^2
+\left[t(1-w)-\frac{\gamma^2}{2}-\gamma_1 r\right]M\pm
\frac{\gamma r}{2}=0,
\end{eqnarray}
where the meaning of the upper and lower sign is the same as
explained just below Eq.~(27).  The difference in the stability
conditions is more pronounced and gives new effects that will be
explained further,
\begin{equation}
\label{eq47} \frac{ (2-w)\gamma M- r -\gamma_1M^2}{1-w} \ge 0,
\end{equation}
\begin{equation}
\label{eq48} \gamma M -wr-w \gamma_1 M^2 \geq 0,
\end{equation}
and
\begin{equation}
\label{eq49} \frac{3(1-w-\gamma_1^2) M^2 + 3 \gamma \gamma_1 M +
t(1-w)-\gamma^2/2 -\gamma_1 r}{1-w}\geq 0.
\end{equation}

The calculations of the phase diagram in ($t,\;r$) parameter space
are done in the same way as in case of  $w=0$ and show that for
$w>0$ there is no qualitative change of the phase diagram.
Quantitatively, the region of first order phase transition widens
both with respect to $t$ and $r$ as illustrated in Fig.~6. On the
contrary, when $w < 0$ the first order phase transition region
becomes more narrow but the condition~(47) limits the stability of
FS  for $r<0$. This is seen from Fig.~7 where FS is stable above
the straight dotted line for $r < 0$ and $t < 0$. So, purely
superconducting (Meissner) phases occur also as ground states
together with FS and FM phases.

\section[]{Conclusion}

We investigated  the M-trigger effect in unconventional
ferromagnetic superconductors. This effect arises from the
$M\psi_1\psi_2$-coupling term in the GL free energy and provokes
the appearance of superconductivity in a domain of the system's
phase diagram that is entirely occupied by the ferromagnetic
phase. The coexistence of unconventional superconductivity and
ferromagnetic order is possible for temperatures above and below
the critical temperature $T_s$, that corresponds to the standard
second-order phase transition
 from normal to Meissner phase -- usual uniform superconductivity
in a zero external magnetic field which occurs outside the domain
of existence of ferromagnetic order. Our investigation is mainly
intended to clarify the thermodynamic behavior at temperatures
$T_s< T < T_f$ where the superconductivity cannot appear without
the mechanism of M-triggering. We describe the possible ordered
phases (FM and FS) in this most interesting temperature interval.

The Cooper pair and crystal anisotropies are investigated and
their main effects on the thermodynamics of the triggered phase of
coexistence is established. Of course, in discussions of concrete
real materials the respective crystal symmetry should be
considered. But the low symmetry and low order (in both  $M$ and
$\psi$) $\gamma$-term in the free energy determines the leading
features of the coexistence phase and the dependence of essential
thermodynamic properties on the type of crystal symmetry is not so
considerable.

Below the superconducting critical temperature $T_s$ a variety of
pure superconducting and mixed phases of coexistence of
superconductivity and ferromagnetism exists and the thermodynamic
behavior at these relatively low temperatures is more complex than
in improper ferroelectrics. The case $T_f < T_s$ also needs a
special investigation.

Our results are referred to the possible uniform superconducting
and ferromagnetic states. Vortex and other nonuniform phases need
a separate study.

The relation of the present investigation to properties of real
ferromagnetic compounds, such as UGe$_2$, URhGe, and ZrZn$_2$, has
been discussed throughout the text. In these compounds the
ferromagnetic critical temperature is much larger than the
superconducting critical temperature $(T_f \gg T_s)$ and that is
why the M-triggering of the spin-triplet superconductivity is very
strong. Moreover, the $\gamma_1$-term is important to stabilize
the FM order up to the absolute zero (0 K), as is in the known
spin-triplet ferromagnetic superconductors.
Ignoring~\cite{Walker:2002}  the symmetry conserving
$\gamma_1$-term does not allow a proper description of the real
substances of this type. More experimental information about the
values of the material parameters ($a_s, a_f, ...$) included in
the free energy (12) is required in order to outline the
thermodynamic behavior and the phase diagram in terms of
thermodynamic parameters $T$ and $P$. In particular, a reliable
knowledge about the dependence of the parameters $a_s$ and $a_f$
on the pressure $P$, the value of the characteristic temperature
$T_s$ and the ratio $a_s/a_f$ at zero temperature are of primary
interest.

\section*{Acknowledgments}
D.I.U. thanks the  hospitality of MPI-PKS (Dresden), where a part
of this work has been made. Financial support
by SCENET (Parma) is also acknowledged.\\

\end{document}